\PassOptionsToPackage{dvipsnames,table}{xcolor}
\documentclass[sigconf]{acmart}

\usepackage{amsmath,amsfonts}
\usepackage{algorithm}
\usepackage{algpseudocode}
\usepackage{graphicx}
\usepackage{textcomp}

\usepackage{mathtools}
\usepackage{nicefrac}

\usepackage{cleveref}
\usepackage{enumitem}
\usepackage{verbatim}
\usepackage{xspace}
\usepackage{siunitx}
\usepackage{soul}
\usepackage{subfig}

\usepackage[font=small]{caption}
\usepackage{mwe}

\usepackage{hyperref}
\usepackage[frozencache,cachedir=.]{minted}

\usepackage{multirow}
\usepackage{makecell}

\newcommand{\ConfMigTool}{CIMig\xspace}

\newcommand{\equivFileTuples}{GHA-Travis CI equivalent set\xspace}

\newcommand{\NewGHAFile}{generated equivalent GHA file\xspace}
\newcommand{\NewTrFile}{generated equivalent Travis CI file\xspace}
\newcommand{\NewCIFile}{generated equivalent Target CI file\xspace}

\newcommand{\FirstPartyTool}{GitHub Actions Importer\xspace}

\newcommand{\NewGHAFiles}{equivalent GHA files\xspace}
\newcommand{\NewTrFiles}{equivalent Travis CI files\xspace}

\newcommand{\RQTwocontent}{How effective is the proposed migration pipeline?\xspace}

\newcommand{\RQThreecontent}{What is the cost of the migration pipeline?\xspace} 
\newcommand{\RQFourcontent}{What are the limitations of our approach
?\xspace}

\newcommand{\review}[1]{\textcolor{black}{#1\xspace}}

\newcommand{\totalJavaProject}{345228\xspace}
\newcommand{\totalTravisOnlyProjects}{13403\xspace}
\newcommand{\totalGithubOnlyProjects}{15888\xspace}
\newcommand{\totalGHAandTravisProjects}{5138\xspace}
\newcommand{\totalGHAandTravisProjectsNoHist}{1748\xspace} 
\newcommand{\totalGHAandTravisProjectsNoSim}{919\xspace} 
 
\newcommand{\totalGHAandTravisProjectsVerfied}{1252\xspace} 
\newcommand{\totalGHAandTravisProjectsTrainSet}{1001\xspace}
\newcommand{\totalGHAandTravisProjectsTestSet}{251\xspace}

\newcommand{\totalTARGeneratedGHA}{2664\xspace}
\newcommand{\totalTARGeneratedTravis}{524\xspace}

\newcommand{\totalNonSimBasedRules}{99586\xspace}

\makeatletter
\def\@copyrightspace{\relax}
\makeatother

\renewcommand\footnotetextcopyrightpermission[1]{} 
\begin{document}
\newcommand{\update}[1]{\textcolor{black}{#1}}
\newcommand{\aseupdate}[1]{\textcolor{black}{#1}\xspace}
\newcommand{\algorithmautorefname}{Algorithm}
\renewcommand{\sectionautorefname}{Section}
\renewcommand{\subsectionautorefname}{Section}
\renewcommand{\subsubsectionautorefname}{Section}
\renewcommand{\paragraphautorefname}{Section}
\renewcommand{\algorithmautorefname}{Algorithm}
\renewcommand{\tableautorefname}{Table}


\title{Example-Based Automatic Migration of Continuous Integration Systems}

\author{Dhia Elhaq Rzig}
\email{dhiarzig@umich.edu}
\affiliation{%
  \institution{University of Michigan Dearborn}
   \country{} 
}

\author{Alaa Houerbi}
\email{houerbi@umich.edu}
\affiliation{%
  \institution{University of Michigan Dearborn}
   \country{} 
}

\author{Chungha Sung}
\authornote{The research work is not related to the author's position in the affiliation.}
\email{chunghs@amazon.com}
\affiliation{%
  \institution{Amazon Web Services}  
   \country{} 
}

\author{Foyzul Hassan}
\email{foyzul@umich.edu}
\affiliation{%
  \institution{University of Michigan Dearborn} 
   \country{} 
}

\bibliographystyle{ACM-Reference-Format}

\begin{abstract}

Continuous Integration (CI) is a widely adopted software engineering practice for faster code change integration and testing.
Developers often migrate \aseupdate{between} CI systems 
\aseupdate{in pursuit of} features like matrix building or easier workflow. However, this migration is effort-intensive and error-prone owing to limited knowledge of the new CI system and its 
syntax. Moreover, our analysis identified that these migrations require multiple iterations and significant time to achieve stability in the new CI system, and there is insufficient support for the automatic migration of CI configurations.

To mitigate this, we propose a novel approach for CI systems' automatic migration: \ConfMigTool. Our approach utilizes example-based mining, where it 
extracts translation rules and configuration patterns from existing migration examples, and employs them to reproduce this migration in new contexts. 
\aseupdate{To empirically validate and evaluate our approach,} we apply it to the migration between Travis CI and GitHub Actions. \aseupdate{We gathered learnings from 1001 projects, and then applied them to migrate} an evaluation set of 251 projects. We also performed a user study employing \ConfMigTool to migrate the CI systems of five Java projects. \aseupdate{These analyses helped us perform a qualitative and quantitative evaluation of  \ConfMigTool, and we contextualize our results by comparing them with those of}
the manual-rule-based \textit{\FirstPartyTool}. Furthermore, our tool generated files that were rated favorably by developers and saved them an average of 42.4 minutes over the manual migration of these same projects.
\aseupdate{Our example learning-based approach is also more flexible,}
\aseupdate{as proven by our ability to apply it to migrate GitHub Actions files to Travis CI}, 
\aseupdate{ which \FirstPartyTool can not do}. We believe \ConfMigTool is the first generic approach of its kind to migrate CI systems and can be applied to other software configuration system migrations.
Our replication package is available at~\cite{repl}.

\end{abstract}

\maketitle
\pagestyle{plain}

\section{Introduction}
\label{sec:introduction}

Continuous Integration (CI) 
is a widely used software engineering (SE) process \aseupdate{for automatically integrating} changes in shared repositories.
It has enabled drastic change and improvement in SE processes and outcomes, such as quicker issue resolution, and faster shipping~\cite{Hilton_2016,zampetti2019study,zhao2017impact}.
Travis CI and GitHub Actions (GHA) are the most popular CI tools for Open Source Software (OSS) projects~\cite{Golzadeh_2022,Hilton_2016,Rzig_2022}, and
\aseupdate{migrations occur frequently between these two tools~\cite{Golzadeh_2022}.}
However, these migrations are  slow and error-prone due to various factors~\cite{Mazrae_2023},
and further complicated by a lack of tool-support. 
\aseupdate{The only 
official tool}, \FirstPartyTool~\cite{gha}, \aseupdate{only supports migrating to GHA}, relies on manual mappings, and lacks support for features \update{such as the migration of secrets like authorization tokens.~\cite{GitHubActionImporter}}. Moreover, this tool is technology-specific and can not be applied to other CI systems.

\aseupdate{Most} existing migration research works focus on analyzing and migrating source code between programming languages~\cite{Java2CSharp, Ramly2006, Mossienko2003, An2018, j2swift}, and few works are concerned with the analysis and migration of configuration code~\cite{Henkel_2020, gligoric2014automated, vassallo2019automated, Rahman2023}, and none tackled the automatic migration of CI configuration code.
 \aseupdate{Many differences exist between source code and configuration.} Source code defines the behavior of software, relies on programming languages like Java, Python, etc., with more descriptive logic syntax, and is generally managed and documented by developers. Configuration code describes the parameters of a software application~\cite{octopus}, relies on markup languages like YAML or domain-specific languages (DSLs)~\cite{ms_cac} with higher abstraction, and is generally maintained by DevOps engineers~\cite{ms_cac}. The migration of CI systems is challenging because of the differences between the Source and Target CI systems~\cite{Mazrae_2023}, owing to the usage of DSLs with higher abstraction. Moreover, our analysis identified that these migrations require multiple iterations and a significant time span to achieve stability in the Target CI system.
To mitigate this difficulty, we propose a novel approach \textit{\ConfMigTool} that employs 
example-based mining, to migrate CI configurations between CI system. \ConfMigTool automatically learns rules from semantically-equivalent tuples of CI files originating from different tools, then applies its learnings to migrate CI files from a  
Source to a Target CI system. To validate and evaluate our approach, we utilized it to perform migrations between GHA and Travis CI.
We assessed
the results of \ConfMigTool through
automatic \review{ and manual} evaluations, described its cost, and analyzed some of the cases where it fails.
Through this paper, we answer the following research questions: 


\noindent \textbf{RQ1:} \RQTwocontent

\noindent \textbf{RQ2:} \RQThreecontent

\noindent \textbf{RQ3:} \RQFourcontent

\ConfMigTool can translate 70.82\% of a Travis CI file and 51.86\% of a GHA file on average.  Its translations \aseupdate{from Travis to GHA} 
 are competitive with \FirstPartyTool, where they had an average cosine~\cite{cosine_sim} similarity of 0.51 to the developer's hand-crafted manual translations, versus 0.45 achieved by \FirstPartyTool. \aseupdate{Unlike the latter,} \ConfMigTool also \aseupdate{translates syntax} in the opposite direction, \aseupdate{where it generates files}
with an average 0.35 cosine similarity to the developer's \aseupdate{versions}.

\noindent Our main contributions through this work are: 

\begin{itemize}[leftmargin=0.25cm,itemindent=0.25cm,labelwidth=\itemindent,labelsep=0cm, align=left,topsep=0cm,noitemsep]
\itemsep0em 
    \item A novel technology-agnostic CI migration technique leveraging Apriori Rule Mining and Tree Association Rules. 

    \item A comprehensive evaluation to evaluate the effectiveness of \ConfMigTool, \aseupdate{and of a few important} failure scenarios. 

    
    \item A dataset of GitHub Actions and Travis CI configuration files from 30,543 real-world Java projects
    \aseupdate{shared at~\cite{repl}}.
    

\end{itemize}

We motivate our work within~\autoref{sec:motivate}. We discuss its background in~\autoref{sec:background}. We detail our approach in~\autoref{sec:approach}, and the quality, cost, and shortcomings of applying techniques to migrations between GHA and Travis CI within~\autoref{sec:results}. Related works are detailed in~\autoref{sec:related_works}, the threats to validity are discussed in~\autoref{sec:threats}, and finally, we conclude our work in~\autoref{sec:conclusion}.

\section{Problem Contextualization}
\label{sec:motivate}
\label{sub:sec:MigrationEffort}

\aseupdate{Prior research~\cite{Mazrae_2023} has identified  through qualitative analysis that migrating a CI infrastructure is a difficult process, due to technical and human hurdles.} 
To further validate these findings, \aseupdate{we performed an empirical study, where we analyzed \aseupdate{1252} projects that migrated from Travis CI to GHA, one of the most common migration patterns~\cite{Golzadeh_2022,Mazrae_2023}. }
These projects were collected through a process detailed in~\autoref{sub:sec:dataset}, and are manually confirmed to have created an equivalent GHA file.\footnote{Defined as sharing 50\% or more functionality, detailed in~\autoref{sub:sec:dataset}}



\aseupdate{Through our Git and API-based analyses,}  we uncovered that an average 
\update{71.20 days, and 2.75} commits are needed to reach a successful build that corresponds \aseupdate{to the equivalent} GHA file, with some projects needing up to 169 commits to reach this threshold. This implies that the migration process is not self-evident and requires multiple attempts over an important span of time.
Furthermore, we find that 48 projects seemingly abandoned the migration process entirely even though they implemented equivalent GHA files, as they never achieved a successful GHA workflow.

\begin{figure}[H]
    \centering
     \includegraphics[width=0.95\linewidth]{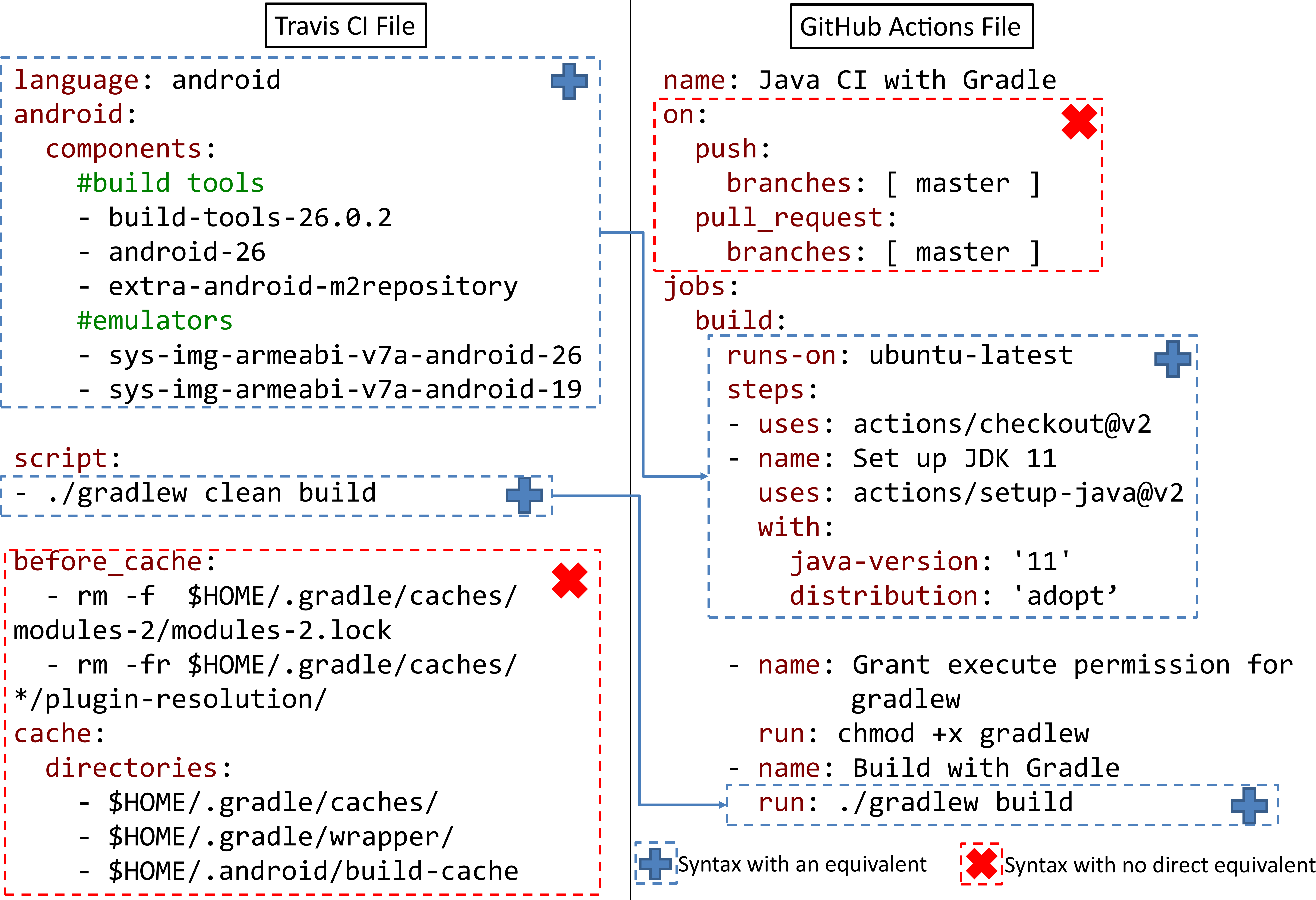}
     \caption{Example of Migration from Travis CI to GitHub Actions}
     \label{fig:motivational_example}
\end{figure}

\aseupdate{To better illustrate the complexity of the CI migration process,}
we present an example of a migration from Travis CI to GHA from the project \texttt{VocableTrainer-Android} in~\autoref{fig:motivational_example}.

The semantically-equivalent segments of the configuration are marked with a \textbf{\textcolor{blue}{+}} sign and linked with an arrow in~\autoref{fig:motivational_example}. However, the sections marked \review{with an}  \textbf{\textcolor{red}{x}} sign, have no direct equivalents between the two syntaxes.




The Travis CI configuration of this project requires specifying \texttt{android} as a language and manually configuring the different \texttt{components} required to run this project within the Travis CI environment.
However, that is not necessary \aseupdate{within GHA}, as all of these components are provided by default \aseupdate{when using} the \texttt{ubuntu-latest} environment.

While Travis CI automatically performs the checkout process and makes Gradle executable,
these steps need to be explicitly performed in GHA. 
Travis CI workflow execution triggers are  configured via its website or performed via API requests~\cite{travis_api}. 
But, GHA developers need to specify them within the \texttt{on} section of the GHA configuration file.
In addition, while Travis CI provides a generic cache configuration mechanism, GHA does not have a workflow-wide directly equivalent syntax for caching. It relies on the configuration of job-specific caches by using \texttt{actions/cache@v3}, or package-manager-specific caching keywords. \texttt{cache:gradle} can be \aseupdate{added to this example} to ensure caching.


\aseupdate{Overall, this example illustrated how  developers need to navigate and avoid many pitfalls during the translation of a CI configuration file, and how the lack of direct equivalents of some syntaxes hinders the translation process}.

\section{Background}
\label{sec:background}
\subsection{Continuous Integration}
Continuous Integration tools automate code integration by automatically validating new commits via the execution of building, testing, and other processes. Most CI tools are configured via Configuration code files, and the execution of a CI tool is referred to as a workflow. 

GitHub Actions~\cite{gha}, Travis CI~\cite{travis}, Azure Pipelines~\cite{azure}, and Circle CI~\cite{circleci} are among the most popular CI tools\review{,} and have \review{many} commonalities. All four tools rely on  YAML-based~\cite{travis_yaml} files to store the configuration of their workflows, with each relying on its own Domain Specific Language (DSL). For all four tools, workflows can be manually or automatically triggered by Git \review{events} such as pull requests or pushes. Workflows are composed of one or more jobs, which may be configured to execute in parallel in different environments. For each of these tools, a job may be composed of one or more steps that run sequentially, and it's possible to use variables to share information between the different steps and the different jobs.

Even with the functional similarity of these tools, there are significant conceptual and syntactical differences between them. For example, while the \review{Operating System} for each GHA job may be specified using the \texttt{runs-on}~\cite{runs} keyword, or the keyword \texttt{vmImage}~\cite{steved0x_2023} for Azure Pipelines, or the \texttt{image}~\cite{oscircleci} keyword for CircleCI,  Travis CI  uses the keyword \texttt{os}~\cite{os} to configure it for all stages and jobs. While Travis CI has a specific phase \texttt{install}~\cite{install} within its lifecycle to prepare the environment, GHA, Azure Pipelines, and CircleCI leave the specification of these phases to the developers. GHA makes workflows and jobs reusable with the keyword \texttt{uses}~\cite{uses}, so does Azure Pipelines with \texttt{task}~\cite{juliakm_2023}, and CircleCI via \texttt{orbs}~\cite{orbs}\review{, but,}
Travis CI does not offer an equivalent function. 


\subsection{\aseupdate{Example-based Learning}}
\label{sub:sub:sec:bg-rule-mining}

\aseupdate{A plausible approach for automatic CI system migration is to learn from how prior developers migrate from source CI systems to target CI systems and how they compose the structure of the CI configurations. Such migration and composition data can be extracted from open-source projects hosted in GitHub. We utilized Association rule mining~\cite{kumbhare2014overview,etemadi2017association}, an ML approach for finding interesting associations among data. Specifically, we used the Apriori Rule Mining~\cite{agrawal1994fast} and  Frequent-Tree Mining~\cite{CMAlg} algorithm to generate rules for the target CI system.}

\subsubsection{Apriori Rule Mining}
Apriori is an Association Rule Mining (ARM) algorithm defined by Agrawal et al.~\cite{agrawal1994fast}. It starts by finding the frequent individual items in a database, also known as transaction set, and expands them to item sets co-occurring together as long as the appearance of those item sets is larger than a minimum threshold specified by the user. Apriori then uses these frequent item sets to generate association rules that reflect general trends in the transactions set. Apriori rules are composed of a Left Hand Side (LHS), the antecedent, also referred to as pre-condition, and a Right Hand Side (RHS), the consequent. Within our work, the transaction set as well as the resulting rules, are composed of subsets of Abstract Syntax Trees (sub-ASTs).


\subsubsection{Frequent Tree Mining and Tree Association Rules}
Frequent Tree mining empowers us to discover frequent maximal, induced, ordered sub-trees with a specific minimum support from a group of similar trees.
We performed Frequent-Tree Mining via the \\  \texttt{CMTreeMiner}~\cite{CMAlg} algorithm on subsets of Abstract Syntax Trees (ASTs).
We grouped these sub-ASTs by their root nodes and passed them as input to \texttt{CMTreeMiner}. Frequent Trees are discussed in detail in Chi et al.'s work~\cite{Chi_2005}. 
Using these trees, we were able to extract Tree Association Rules (TAR), which we adapted from the work of Mazuran et al.~\cite{Mazuran_2009}.
Similar to association rules, TARs are composed of an antecedent and a consequent. 
Within our work, we considered the antecedent to the root node as well as 50\% of the branches of a Frequent Trees,  and the consequent being the remaining branches of the tree. Hence, a Frequent Tree may generate multiple TARs during the execution of \ConfMigTool. 

\section{Approach}
\label{sec:approach}

 \begin{figure*}[htbp]
    \centering
    \includegraphics[width=0.9\linewidth ]{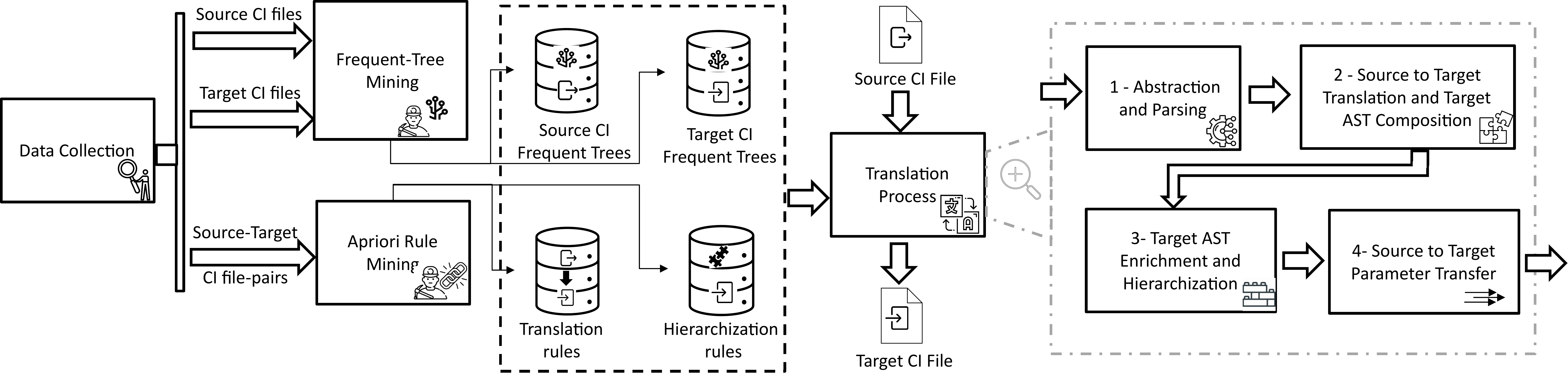}
    \caption{Overview of \ConfMigTool when used to migrate between Travis CI and GHA}
    \label{fig:all_overview}
\end{figure*}
An overview of our \aseupdate{approach} is shown in~\autoref{fig:all_overview}. 
We use Travis CI and GHA syntaxes in the different illustrative examples. 

\subsection{Data Preparation}
\label{sub:sec:dataset}

The proposed approach \ConfMigTool requires three sets of configuration files. Set (1) containing Source CI files , Set (2) Target CI files, and Set (3) of Source and Target CI file-pairs, with each pair containing two files from different CI tools that implement similar functionality. To prepare these files for the analyses we aim to perform on them, we apply an abstraction process to them. This process parses these files into equivalent ASTs, then transforms their leaves by matching them with regular expressions that contain predefined keywords to preserve the commands used within the configuration code files while removing their project-specific parameters. \aseupdate{To evaluate our approach,} we chose to focus on Java projects using Travis CI or GHA as they are the most important subset of CI-using OSS projects~\cite{Hilton_2016,Beller_2017,Durieux_2019,Rzig_2022,Rzig_Hassan_2022}. \aseupdate{ To concertize the data preparation phase in this context, we discuss the processes we followed to create the 3 aforementioned sets in this specific context.} First, we collected the projects from two sources: Google BigQuery and GitHub, the two most popular OSS repository hosting sites~\cite{Lisowski_2021,github_website}, after applying criteria on activity and popularity as outlined by previous works~\cite{Munaiah_2017,Kalliamvakou_2016,Gousios_2017}, ensuring that these projects have a size $>$ 0 KB, have been active in 2021, and have a popularity $\ge 5$ stars or $\ge 5$ forks. We collected \totalJavaProject projects after de-duplication. Then, we used Travis CI and GHA APIs to establish a project's usage of these CI tools, a more accurate method of establishing adoption~\cite{Rzig_2022}.

This allowed us to build these three project sets:

\begin{itemize}[noitemsep,topsep=0pt]
    \item Travis CI-Only projects: \totalTravisOnlyProjects, \aseupdate{containing Set (1) or Set (2), depending on Translation direction.}
    \item  GHA-only projects: \totalGithubOnlyProjects, \aseupdate{containing Set (1) or Set (2), depending on Translation direction. }
    \item Travis CI and GHA projects: \totalGHAandTravisProjects, \update{\totalGHAandTravisProjectsVerfied after filtering, \aseupdate{ containing Set (3).}}
\end{itemize}

\aseupdate{We used the first two projects sets \aseupdate{to extract Set (1) and Set (2)} for Task B in~\ref{sub:sub:sec:training}, to extract Frequent Trees for both Travis CI and GHA.} 
We used the third project set to perform migration effort analysis discussed in~\autoref{sec:motivate}, and \aseupdate{to extract Set (3)} of semantically equivalent configuration code file tuples for Tasks A-1 and A-2 in~\ref{sub:sub:sec:training}. \review{It’s important to note that while Travis CI uses one configuration code file, GHA may use multiple files, hence why we're extracting tuples from a project, as they contain one Travis file, and may contain more than one GHA file.} We applied the following \update{filtering} process to find these tuples.
First,  we performed a git-history-based analysis to extract the  Travis CI and GHA file tuple \review{which contains the file pair composed of the Travis CI and one of the GHA files} with the highest cosine similarity~\cite{cosine_sim}, a metric used within previous works~\cite{Chen_Monperrus_2019,Xie_2020} to determine source code and configuration code similarity,
\review{and which have passing build statuses as confirmed by the GHA and Travis CI APIs.} We eliminated \totalGHAandTravisProjectsNoHist projects as they did not have configuration files for one or both tools in their history, likely due to Git rewritings~\cite{Bird_2009}, and
\totalGHAandTravisProjectsNoSim projects, due to their tuples having a maximum cosine similarity of 0.1 or less. 

Second, to confirm the semantic equivalence of the remaining tuples from the third set, two co-authors manually analyzed file tuples from 2471 projects. As mentioned earlier, the extracted file tuple may contain more than one GHA file. Hence, the developers performed a pairwise comparison between the Travis file and each of the GHA files in each tuple, starting with the longest GHA file. To save time, they stopped when reaching the minimum equivalency criterion. We opted for this permissive semantic equivalence criterion after perceiving that very few projects completely re-implement the same functionality between GHA and Travis CI, which is consistent with previous findings~\cite{Mazrae_2023}. After applying these filtering processes, only \totalGHAandTravisProjectsVerfied projects met these criteria. 

Similar to other works that tackled code translation~\cite{Roziere,ahmad-2021}, we split third set into two subsets, \update{following the 80\%-20\% ratio}, a "training" set of file tuples from \totalGHAandTravisProjectsTrainSet projects and a "test" set of file tuples from \totalGHAandTravisProjectsTestSet projects. The project splitting process was random to maintain representativeness. \aseupdate{Only the training set is used for Tasks A-1 and A2, while the testing set is reserved for the evaluation of the approach}. \aseupdate{Since Tasks A-1 and A2 of \ConfMigTool are designed to learn on file-pairs, we transform each tuple into pairs where the same Travis CI file is paired with the multiple GHA files}.


\label{sub:sec:approach}
\subsection{Training \ConfMigTool}
\label{sub:sub:sec:training}
\subsubsection{\textbf{\review{Task A:} Apriori Rule Mining Process}}
\label{sub:sec:rule_mining}
The goal of this process is to find rules that allow us to translate \aseupdate{Source CI syntax to Target CI Syntax}, which we refer to as  \review{\textcircled{1}} Translation Rule Mining as well as rules that link different parts of \aseupdate{Source CI syntax to each other}, referred to as \review{\textcircled{2}} Hierarchization Rule Mining. \aseupdate{This process is applied to the Source-Target CI file pairs set.}

\begin{figure}[htbp]
    \includegraphics[width=.9\linewidth ]{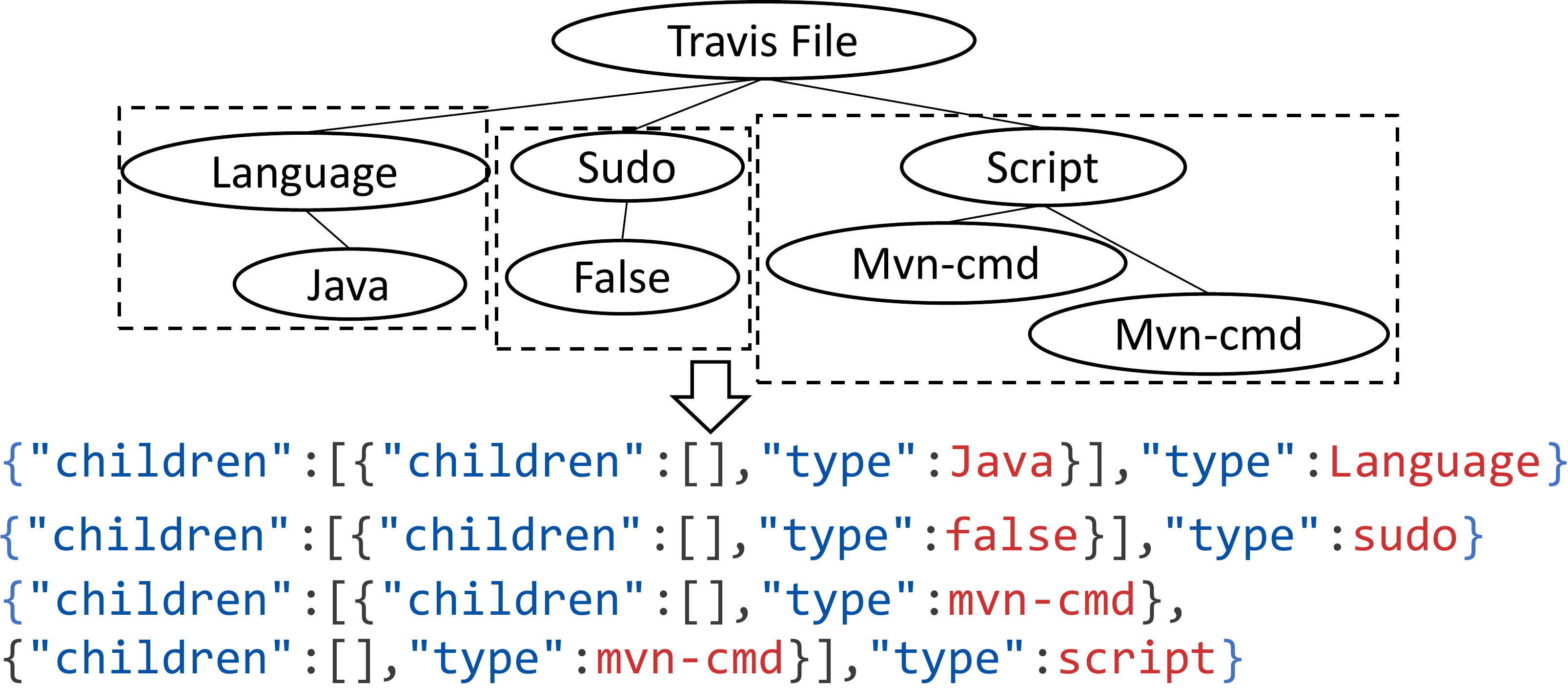}
    \caption{Travis CI H-2 AST Extraction Example}
    \label{fig:h2_collect_example}
\end{figure}

\noindent\textbf{\review{Task A-1: }Translation Rule Mining.} 
To extract translation rules to guide our translation \aseupdate{from the Source CI to the Target CI tool, }
we analyze the previously-prepared file pairs.

First, after the abstraction of these files as detailed within~\autoref{sub:sec:dataset}, we parse them into ASTs. Then, for each pair of ASTs, we extract the sub-ASTs of height equal to 2 starting from the leaves of the ASTs, which we refer to as H-2 ASTs within this work, and we represent them in a textual format. We decided on this height after a process of parameter tuning, \review{detailed in the Parameter Tuning paragraph of ~\ref{par:parameter_tuning}.} An example of the application of the abstraction and H-2 collection processes is shown in~\autoref{fig:h2_collect_example}. \aseupdate{For each file pair $i$, we obtain a set of H-2 ASTs extracted from the Source CI file: \( {SRC-H2}_i\), and a  set of H-2 ASTs extracted from the Target  CI file \( TGT-H2_i\) }

Second, for each file pair, we apply Cartesian product, used in other code-translation works~\cite{Teyton_2013}, to create a transaction set \aseupdate{ $ T_i = SRC-H2_i \bigtimes TGT-H2_i $}
We chose  this product since the alignment of the configuration code files \aseupdate{from different tools} is not possible in many of cases, as different configuration parameters can be at different locations within the file pairs
due to some tools, such as GHA, \review{employing} a more flexible file structure than others.

Third, all the transaction sets generated from the file pairs are grouped into one large set \aseupdate{ $ T_{transl} = \sum_{i=1}^{N} T_i $ }
on which we perform our Apriori-based Association Rule Mining (ARM)~\cite{agrawal1994fast}, previously detailed in~\autoref{sub:sub:sec:bg-rule-mining}.  This rule-mining was used in other works tackling code translation and configuration mapping such as \review{that} of Hora et al.~\cite{Hora_2015}

These rules are evaluated in terms of their support~\cite{agrawal1994fast}, confidence~\cite{agrawal1994fast}, and lift~\cite{mcnicholas2008}, with higher values indicated higher quality rules.
Support reflects how often the item set appears together,
$support(SRC \Rightarrow TGT )=P( SRC \cup TGT )$,
where $SRC$ is a specific H-2 AST from Source CI, $TGT$ is a specific H-2 AST from Target CI.
Confidence reflects how often the rule is correct,
$confidence( SRC \Rightarrow TGT  ) =\nicefrac{ P( SRC \cup TGT)}{P( SRC )}$.
Lift is the ratio of the actual confidence of a rule to its expected confidence, 
$lift(SRC \Rightarrow TGT ) = \nicefrac{confidence(SRC \Rightarrow TGT )}{P( TGT )}  $.
We specified a minimum support of $10^{-6}$, a value determined via a process \review{detailed in the Parameter Tuning paragraph of ~\ref{par:parameter_tuning}.}

Fourth, we filter the generated rules to keep those with the format of \update{SRC-CI-H2-AST $\Rightarrow$ TGT-CI-H2-AST,} thus creating the Rule-Set \(R\).

We calculate the confidence, lift and support products of these rules with their flipped counterparts, of the format \update{ TGT-H2-AST $\Rightarrow$ SRC-CI-H2-AST}, to substantiate the equivalence between the two H-2 ASTs. \update{This is important since} one Source CI H-2 AST may have multiple possible equivalent Target H-2 ASTs, and vice-versa.

Finally,
 we automatically bifurcate \(R\) into two sets, based on whether the cosine similarity of the LHS's leaves and the RHS's leaves was above 0.5. These sets are:

\noindent \textit{\aseupdate{\(R_{sim}\)}:}  Similarity-Based \review{rule-set}.


\noindent\textit{\aseupdate{\(R_{stat}\)}:} 
 Statistical-Based \review{rule-set}.

\aseupdate{We opted for this bifurcation as we anticipate a number of spurious rules will be present in \(R\) due to the application of the Cartesian product. We choose to apply \(R_{stat}\) in the translation process,  after applying  additional constraints, as it may still contain some useful non-textually-similar rules. We detail these constraints and the translation process in~\autoref{sub:sec:TranslationProcess}.}
\newline
\noindent\textbf{\review{Task A-2: }Hierarchization Rule Mining.}
The translation rules only capture two levels of the entire Source CI AST to find its equivalent Target CI AST. However, CI ASTs often contain 3 or more levels. Thus, multiple H-2 ASTs can be linked with a variety of intermediate nodes on multiple levels.
To better infer the intermediate nodes within a Target AST, and ensure the correct \aseupdate{composition of the generated Target CI file}, we created a set of hierarchization rules via the following steps.

First, from each Target CI file \(i\), we built a transaction set \(TH_i\) of the extracted H-2 AST nodes and their parents, then we built \(TH = \sum_{i=1}^{N} TH_i \)
on which we ran the Apriori algorithm with a minimum support of $10^{-6}$ to generate the hierarchization rules.  

\update{
Similar to their translation counterparts, the rules were filtered to keep those of the desired format of H-2-AST-Child $\Rightarrow$ Parent.
These rules allow us to find the direct parents of an H-2 AST, thus allowing us to infer some intermediate nodes within the complete generated AST of our \aseupdate{Generated Target CI file}. In addition, their confidence, lift, and support products were also calculated.}

\subsubsection{\textbf{\review{Task B:} Frequent-Tree Mining Process.}}
\label{sub:sec:frequent-tree-mining}
While our translation and hierarchization rules allow us to translate \update{H-2} ASTs \update{and find their direct ancestors, they do not capture patterns that link multiple H-2 CI ASTs to each other or patterns that span more than 2 levels.}
\update{Such patterns may allow us to add beneficial sub-ASTs via inferring which sub-ASTs occur together, thus allowing us to address syntax that is not directly translatable from the Source CI syntax or syntax that does not have a direct equivalent.}
To capture these useful patterns, we perform Frequent-Tree Mining, detailed in~\autoref{sub:sub:sec:bg-rule-mining},
on the \aseupdate{Source CI files set, and the Target CI files set}. After abstracting these files, we extract sub-ASTs originating from each of their intermediate nodes.
This mining process empowered us to discover a set of Frequent sub-ASTs, \aseupdate{which we refer to as \(FT\), where each tree has }
a minimum support of 5\%, a value we chose via parameter tuning\review{, detailed in the Parameter Tuning paragraph of ~\ref{par:parameter_tuning}}.  
\begin{figure}[htbp]
    \centering
    \includegraphics[width=0.8\linewidth ]{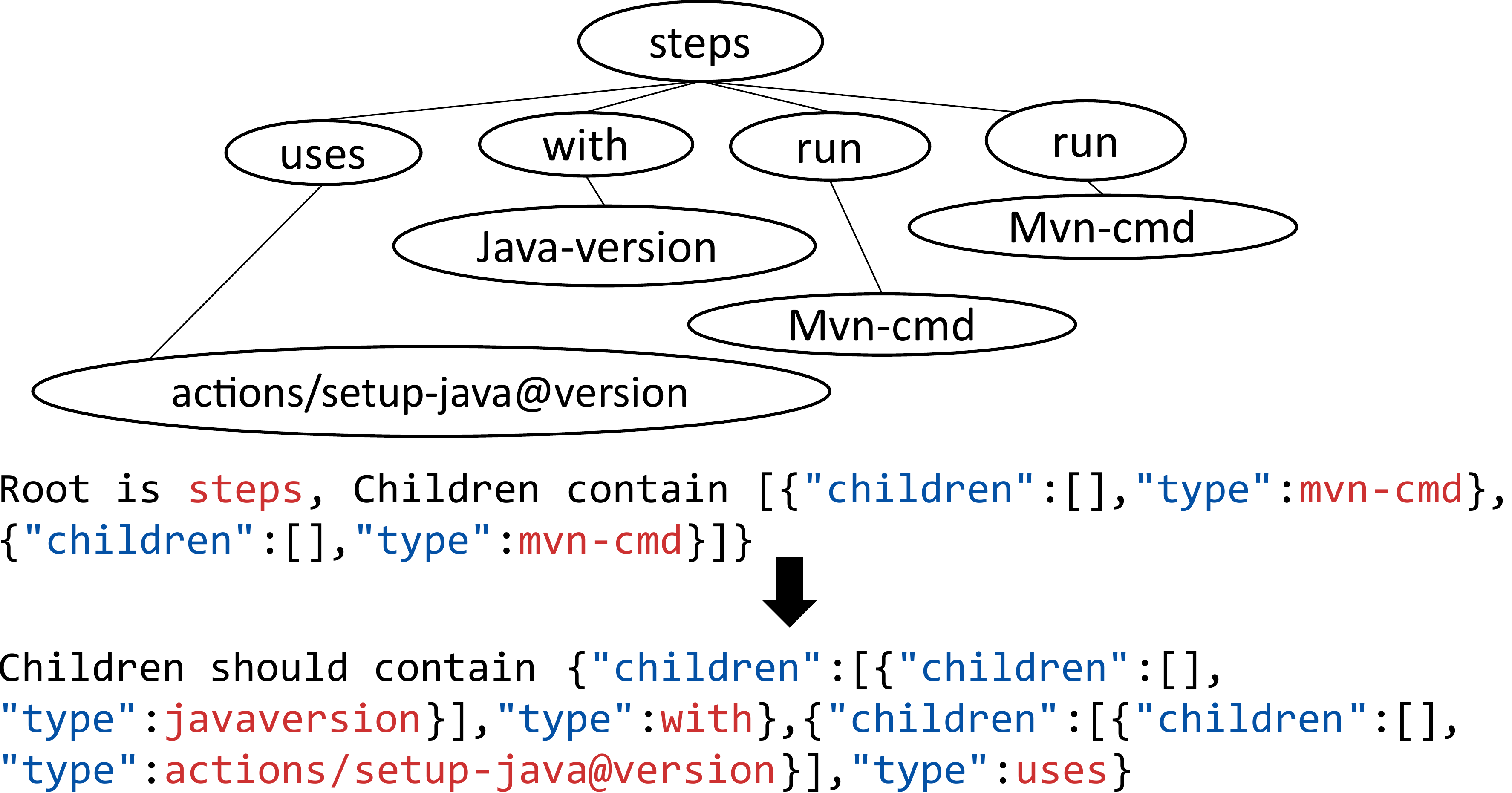}
    \caption{Example of Frequent Tree mined from GHA and Generated TAR}
    \label{fig:travis_tar}
\end{figure}

These Frequent Trees we extracted capture sub-ASTs 
which co-occur frequently within the files we used as input, and 
an example of such a Frequent Tree \update{containing a beneficial pattern}, which we extracted by mining GHA \aseupdate{files}, is shown within~\autoref{fig:travis_tar}. This tree contains the syntax used to setup and configure Java within a specific job, as well as the usage of Maven commands, signaling that these two elements are likely to occur together. As detailed in ~\autoref{sub:sub:sec:bg-rule-mining}, these Frequent Trees generate multiple Tree Association Rules (TAR), and an example TAR is shown in the figure as well,
where the antecedent is the root node steps along with the usage of Maven commands within an AST, and the consequent is the setup and configuration of Java. \aseupdate{Such co-occurring H2-ASTs can't be be identified by translation and hierarchization rules.}  \update{ As illustrated by this example, the configuration of Java is a beneficial addition to our translation. 
Furthermore, TARs generally add more intermediate nodes to an AST file, which are useful for the hierarchization process.}

\subsubsection{\textbf{\review{Parameter Tuning for Task A \& B}}}
\label{par:parameter_tuning}
\review{
While designing the Apriori Rule Mining and Frequent Tree Mining processes, we followed an extensive parameter tuning process. \aseupdate{Due to the prevalence of the migration from Travis CI to GHA, we focused on that translation scenario during this process. 
}}
\review{
For the Apriori Rule Mining tasks, while deciding on the optimal number of levels to capture within our translation rules, our goal was to generate rules that strike a good balance between conservativeness and generality, as higher-order rules may less easily accommodate the migration with new CI workflow steps not seen during the training process, while lower order rules may generate too many rules that are prone to noisiness. \aseupdate{To determine the ideal number of levels to consider, we mined rules with different sub-ASTs of different heights.}  
We specifically evaluated 3 different types of rules: H-2 rules, with two levels on both sides of the rules,
Mixed rules, with three levels on one side, two levels on the other side of the rule, 
and H-3 rules, with three levels on both sides of the rules.
For each type of rule, we mined Travis CI => GitHub Actions rules, and then performed an evaluation of these rule sets against hand-crafted ones.
While Sim-based H-2  and Stat-based H-2 rules had F-1 scores of 71.25\% and 31.85\%, Sim-based Mixed and Stat-based Mixed rules had F-1 scores of 60.75\% and 8.10\%, and Sim-based H-3  and Stat-based H-3 rules had F-1 scores of 33.33\% and 10.26\%. Hence, it's clear that H-2 rule-set has significantly better rules, while considering higher-level rules causes a precipitous drop in rule quality. Furthermore, while performing the rule mining process, we experimented with different values for the minimum support, and opted to use $10^{-6}$ as it allows the generation of the maximum number of rules on the development machine, described in~\ref{sub:sec:approach-cost}, without causing memory consumption issues related to the Apriori algorithm~\cite{agrawal1994fast}}

\review{ Concerning Frequent Tree Mining, 
we again aimed to strike a balance between conservativeness and generality, and to operate within the constraints of time and memory needed for the mining process. Hence, we attempted the mining process with multiple minimum support values ranging from 1\% to 75\%. 
Of the values in this range, we found that a minimum support of 5\% generated a sufficient number of trees, consisting of 2664 GHA Frequent Trees and 524 Travis CI Frequent Trees, within an amount of time detailed in \ref{sub:sec:approach-cost}, while higher support values resulted in a much smaller number of Frequent-trees, especially for Travis CI. For example, 10\% minimum support resulted in the discovery of only 1006 GHA trees and 175 Travis CI trees, and 25\% resulted in the discovery of only 191 GHA trees and 40 Travis CI trees, thus capturing far fewer patterns. Frequent Tree Mining with minimum support values lower than 5\% either went on indefinitely, or took much longer time and resulted in few additional trees, most of which were not generalizable.}

\subsection{Using \ConfMigTool}
\label{sub:sec:TranslationProcess}


The four steps of the approach that we follow to translate files from the Source CI syntax to the Target CI syntax are illustrated in~\autoref{fig:all_overview}.
Within this section, we detail the different steps of the translation process and illustrate them with an example of a translation from Travis CI to GHA.
\review{Similar to \FirstPartyTool, we designed \ConfMigTool to use one file as input and produce one file as output, as the splitting of CI configuration across all files is optional in some CI tools such as GHA, and not at all supported by other tools such as Travis CI.}



\begin{figure}[H]
    \centering
    \includegraphics[width=0.75\linewidth]{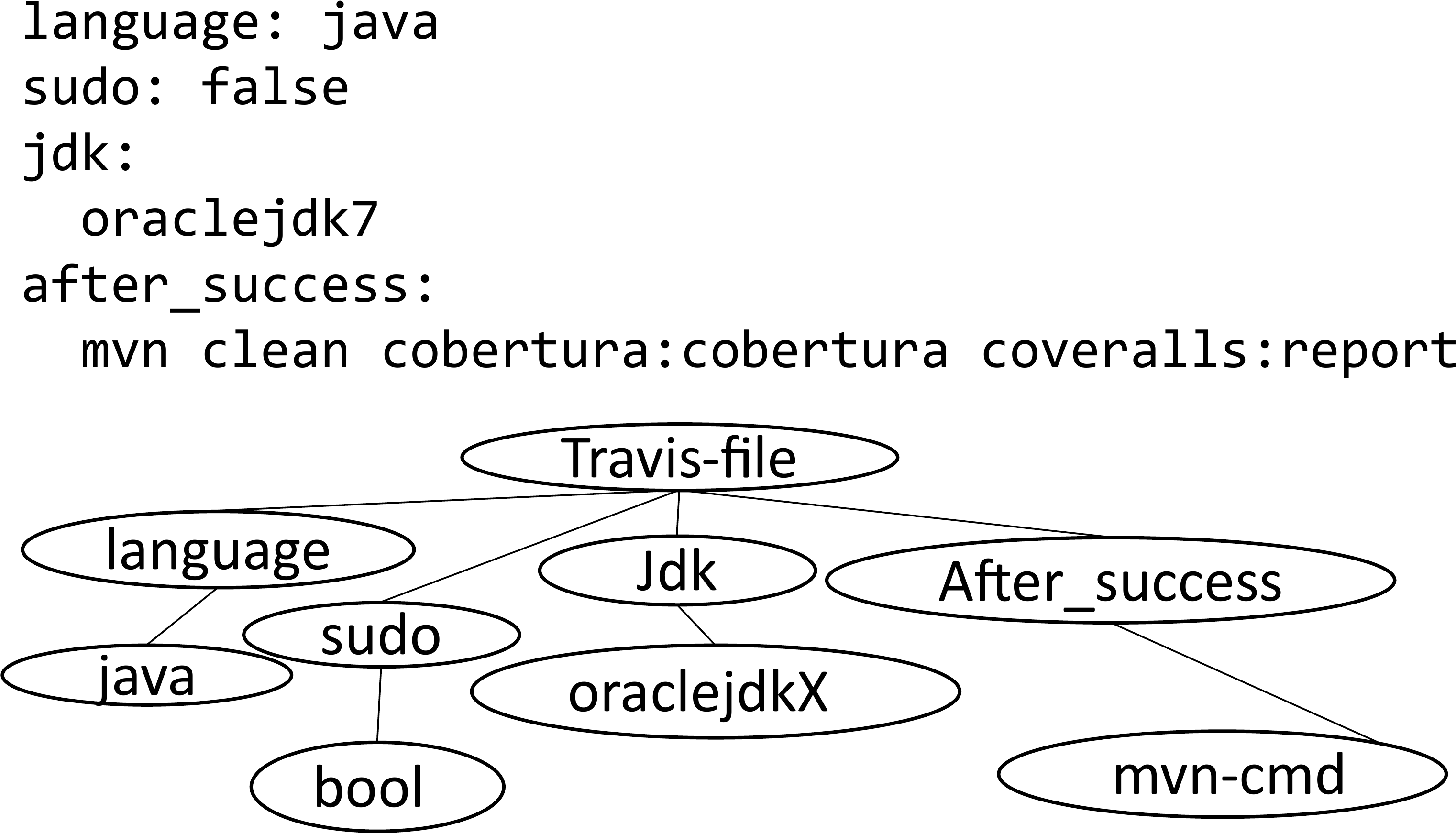}
    \caption{Example of Travis CI File and its corresponding AST}
    \label{fig:step1}
\end{figure}

\subsubsection{Step 1: Abstraction and Parsing}
\label{sub:sub:sec:parsing}

First, the configuration code of the source file is processed with the same abstraction process applied during the training phase, described in ~\autoref{sub:sec:dataset}, and then parsed to an Abstract Syntax Tree (AST) from which we collect the H-2 ASTs. 
The parameters of the commands within these nodes, which are removed in the abstraction process, are stored for usage in a later step. An example of a Travis CI configuration file and its equivalent abstracted AST is shown within~\autoref{fig:step1}.

\begin{figure}[htbp]
    \centering
    \includegraphics[width=\linewidth ]{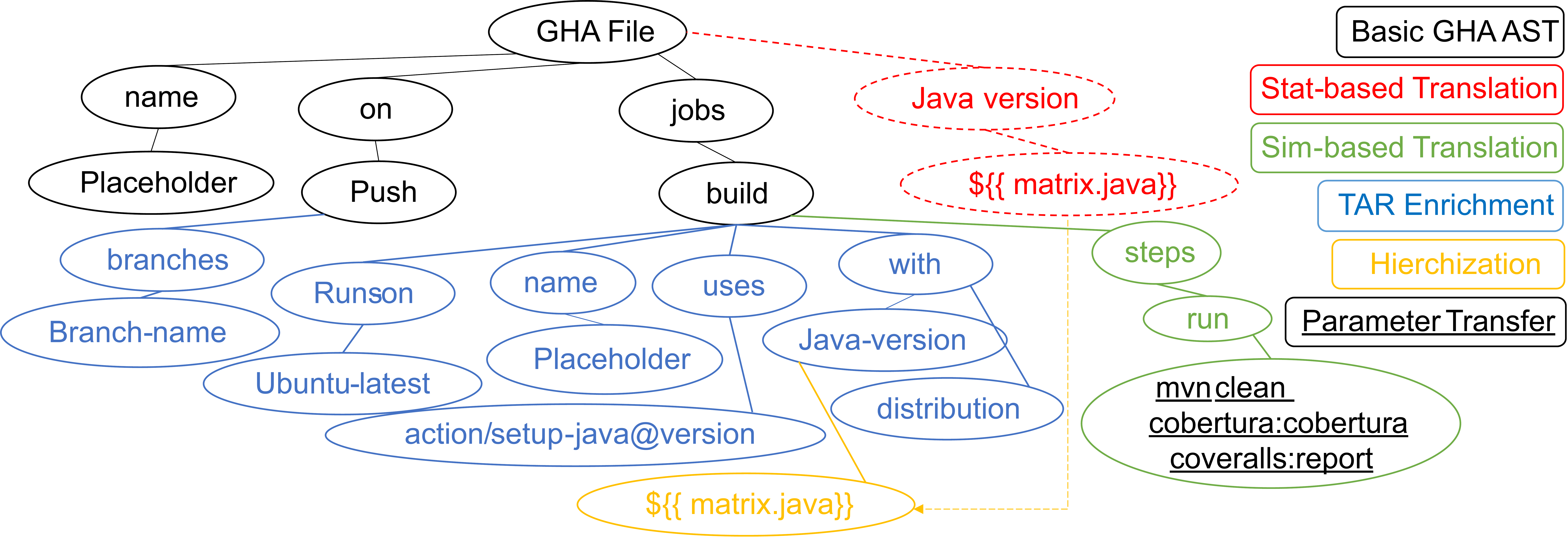}
    \caption{Example of a translated GHA AST}
    \label{fig:steps}
\end{figure}

\subsubsection{Step 2: Source to Target Translation and Target AST Composition}
\label{sub:sub:sec:ast_comp}
\update{
This step is composed of 3 phases: Initialization, Sim-based Translation, and Stat-based Translation. We also detail the Insertion Process we follow during the latter two phases. }

\noindent\review{Step 2.1: }\textbf{Initialization.}
First, we initialize a \update{Target CI} \textit{seed tree} before the translation process begins. This AST \update{is created from Frequent Trees found within the Target CI files, and that was verified to follow the correct structure of the Target CI tool.} 
An example of a basic GHA AST \update{composed of a seed tree} is shown in black in~\autoref{fig:steps}.
This AST forms the basis of the file we're attempting to create as an end result of our translation process, and we refer to this file as \update{\NewCIFile.} 

    \begin{figure}[htbp]
    \includegraphics[width=0.9\linewidth]{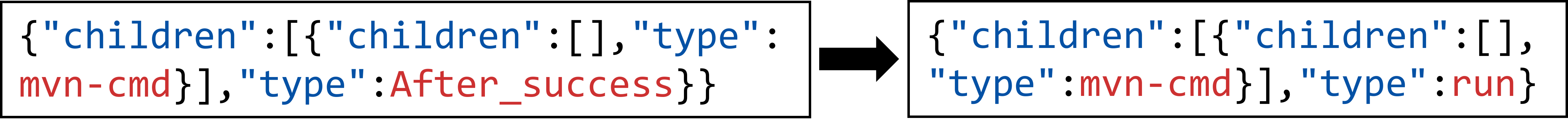}
    \caption{Example of Sim-based translation rule}
    \label{fig:step2a}
        \end{figure}

\noindent\review{Step 2.2: }\textbf{Sim-based Translation.}
Second, we attempt a Sim-based translation, which makes use of Sim-based rules, detailed in \review{Task A-1 of}~\autoref{sub:sec:rule_mining}. For each \update{Source CI} H-2 AST \update{collected} within the previous step, we collect all the Sim-based rules with an LHS that matches it. Then, we extract the best rule according to its confidence product and apply it to generate the corresponding \update{Target CI} H-2 AST, which is then inserted within the AST of the generated equivalent Target CI file.
\autoref{fig:step2a} shows an example of such a translation rule that applies to the Travis CI file shown in~\autoref{fig:step1}. The usage of its results in a \NewGHAFile is shown in \textcolor{Green}{green} in~\autoref{fig:steps}. These rules effectively translate syntax that is directly equivalent between Source and Target CI systems and has textual similarity.

\begin{figure}[htbp]
\includegraphics[width=0.9\linewidth]{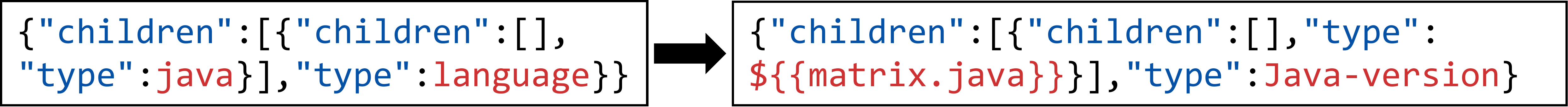}
    \caption{Example of Stat-based translation rule}
    \label{fig:step2b}
\end{figure}

\noindent\review{Step 2.3: }\textbf{Stat-based Translation.}
Third, we attempt Stat-based translation. 
For each H-2 AST not translated within the previous step, we collect all the Stat-based rules with an LHS that matches it. However, we look for certain prerequisites before attempting to apply the Stat-based rules. For each rule, we collected the Frequent Trees of \update{ the Source CI tool}, which contain the LHS of this rule, and the \update{Target CI } Frequent Trees, which contain the RHS of this rule.\footnote{This step is independent of the translation process, it is pre-computed to help accelerate it.} \update{\ConfMigTool analyzes each matched rule in descending order of their confidence product. It ascertains whether \update{at least one  Source CI} Frequent-Tree, containing the LHS of the Stat-based rule, is present within the Source CI file's AST. It also verifies if a Frequent-tree \update{ from the Target CI tool}, containing the RHS of the Stat-based rule,
\review{has} at least 50\% of \review{its branches} within this AST}.
If both conditions are met, the rule is applied and the corresponding \update{Target CI} H-2 AST is generated and inserted within the AST of the \update{\NewCIFile}. \autoref{fig:step2b} shows an example of such a translation rule that applies to the Travis CI file shown in~\autoref{fig:step1}. The usage of its results in a \NewGHAFile is shown in \textcolor{Red}{red} in~\autoref{fig:steps}. This type of rule is especially useful for the non-directly-equivalent syntax and directly-equivalent syntax which does not have textually similar leaves, such as the translation of the \texttt{language:android} segment from the motivational example in~\autoref{fig:motivational_example}.

\noindent\textbf{Insertion Process.}
\update{In this paragraph, we detail} the insertion process \update{that we followed during the Sim-based Translation and the Stat-based Translation}. 
\ConfMigTool performs a DFS-based search within the \update{\NewCIFile} AST to find the deepest node that matches the parent node of the new H-2 AST, which is then used as the point of insertion. The H-2 AST's \update{children} are inserted as the matching node's siblings.
The design of this process was guided by observations of the YAML \aseupdate{syntax, which is used by many CI tools}, as intermediary nodes do not occur multiple times on the same level within a YAML file. If no matching nodes are found, the new H-2 AST is assumed to be a direct descendant of the file's root node and is accordingly inserted at the root of the file.

\subsubsection{Step 3: Target AST Enrichment and Hierarchization}~\\
\label{sub:sub:sec:file_inter_nodes}

\noindent \update{
\review{Step 3.1: }\textbf{AST Enrichment with TARs.}
to improve the structure of our \NewCIFile, we make use of TARs contained within the previously-mined Frequent Trees, detailed in \review{Task B of}~\autoref{sub:sec:rule_mining}. TARs can add beneficial patterns found within CI files of the same type, as well as intermediate nodes and structures that can be used to hierarchize the previously generated H-2 ASTs.} 
We attempt to match each of the Target CI tool's TARs with the AST of the \NewCIFile. If a TAR is applicable,
we insert the AST branches it generates while preserving any existing nodes within the file.
\update{
an example of an AST Enrichment is shown in \textcolor{Blue}{blue} in~\autoref{fig:steps} }.


\noindent\review{Step 3.2: }\update{\textbf{AST Hierarchization.}}
\update{The goal of the hierarchization process is to improve the placement of our H-2 ASTs, and the internal structure of our \NewCIFile's AST.} We apply~\autoref{alg:hierarchization} to achieve this process, \update{which employs the hierarchization rules, detailed \review{Task A-2 of} in~\autoref{sub:sec:rule_mining}.} First, \update{as shown in lines 12-13}, for each H-2-AST we inserted at the root, we attempt to apply the hierarchization process. \update{ Within this paragraph, we refer to the H-2 AST we're attempting to hierarchize as the current H-2 AST.}
\update{For each current H-2 AST, we call the function DFS\_Based\_Insert, detailed in lines 1-11,
where we perform a DFS-based search to find the deepest node that matches the current H-2 AST's parent type. If a match is found, we
insert the current H-2 AST's children as children of the matching node and remove the current H-2 AST from the root node.} This re-application of the same insertion process we followed in the previously-described \textit{Step 2} allows us to take advantage of the new \update{intermediate} nodes added via the TAR enrichment process that \update{we previously} applied.
\update{ If no matches are found,} 
we collect all the \update{Target CI} hierarchization rules the LHS of which matches the current H-2 AST and we apply the \update{hierarchical} rule with the highest confidence product, \update{as detailed in lines 14-21.}
\update{ Lines 22-30, show how we use this rule: we} produce a new node using the new parent type, and add the current H-2 AST to its children.
\update{We then pass this new node as a search target to DFS\_Based\_Insert. If a node with the same type as our new parent is found within our Target CI file AST, we insert the current H-2 AST as one of its children}. If no matches are found, the \update{ newly generated node is inserted
a child of the root of the \NewCIFile's AST.}



\begin{figure}[htbp]
    \includegraphics[width=0.8\linewidth]{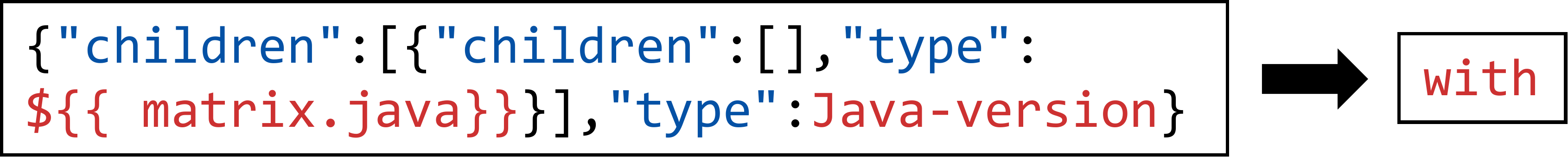}
    \caption{Example of Hierarchization rule}
    \label{fig:step4}
\end{figure}

An example of a hierarchization rule is shown in~\autoref{fig:step4}. It applies to \update{the \NewGHAFile we're constructing in~\autoref{fig:steps}}, where the usage of this rule's results is shown in \textcolor{Dandelion}{yellow}. \update{This example also illustrates how the application of TARs allowed us to add new intermediate nodes, which were then useful during the hierarchization process.}

\subsubsection{Step 4: Source to Target AST Parameter Transfer}
\label{sub:sub:sec:param_transfer}

Before applying the abstraction process in~\autoref{sub:sub:sec:parsing}, we stored the parameters that correspond to the different commands contained in the H-2 ASTs we extracted.
Throughout the different steps of our translation process, we keep track of which parameters correspond to each collected H-2 \update{Source CI} AST, as well as which generated H-2 \update{Target CI} AST corresponds to which  H-2 \update{Source CI} AST. The generated \update{Target CI} ASTs are abstract due to the nature of the rule generation process, making the parameter transfer to them a direct process, where we copy the parameters to the new commands while preserving their order. Hence, we end up with a \NewCIFile that contains commands identical to their Source CI counterpart. An example of the results of this step is illustrated within
the underlined node in the AST shown in~\autoref{fig:steps}, where the parameters of the maven command were transferred from the original Travis CI AST. The \NewCIFile AST is finally transformed into a regular YAML file that can be used by the developers in their \update{Target CI} environment.

\setlength{\textfloatsep}{0.2cm}
\begin{algorithm}[htbp]
    \begin{algorithmic}[1]
    \scriptsize
        \caption{Hierarchization Algorithm}
        \label{alg:hierarchization}
        \Function{DFS\_Based\_Insert}{$CI\_H2\_AST$,$CI\_AST$} 
            \State $T \gets CI\_H2\_AST.Parent\_Node.Type$
            \State $ Insert\_Node \gets$ \Call{DFS Based Search}{$CI\_AST,T$} 
            \If{$ Insert\_Node \neq NULL$} 
                \State$Insert\_Node.Children \gets (Insert\_Node.Children\_AST \cup  CI\_H2\_AST.Children)$
                
                \State  $CI\_AST.Children \gets  (CI\_AST.Children \setminus   CI\_H2\_AST)$
                
                \State \Return True
            \Else
                \State \Return False
            \EndIf
        \EndFunction
        
        \ForAll{$CI\_H2\_AST \in CI\_AST.Children$ } 
        
            \If{\Call{DFS\_Based\_Insert}{$CI\_H2\_AST$,$CI\_AST$} = False}

                
                \State $Matched\_Hierarch\_Rules \gets \emptyset$  
                \ForAll{$Hierarchy\_Rule \in Hierarchy\_Rules$  } 
                    \If {$Hierarchy\_Rule.LHS = CI\_H2\_AST$} 
                        \State $Matched\_Hierarch\_Rules \gets (Matched\_Hierarch\_Rules \cup Hierarchy\_Rule)$
                    \EndIf
                \EndFor
                \If {$Matched\_Hierarch\_Rules.size > 0 $}
                    \State $Best\_Rule \gets Best(Matched\_Hierarch\_Rules)$
                    \State $ New\_CI\_AST \gets $ \Call{Init}{$Best\_ Rule.Parent\_Node.Type$}
                    \State $ New\_CI\_AST.Children \gets (New\_CI\_AST.Children\cup CI\_H2\_AST)$
                    \State $ CI\_AST.Children \gets (CI\_AST.Children \setminus CI\_H2\_AST)$
                        \If {\Call{DFS\_Based\_Insert}{$New\_CI\_AST$,$CI\_AST$} = False} 
                            \State $ CI\_AST.Children \gets (CI\_AST.Children \cup New\_CI\_AST)$  
                        \EndIf
                \EndIf
            \EndIf
        \EndFor      
    \end{algorithmic}
\end{algorithm}


\section{Evaluation}
\subsection{\aseupdate{RQ1: How effective is \ConfMigTool?}}
\label{sub:sec:accuracy_eval}
\aseupdate{To measure the effectiveness of \ConfMigTool, we performed two-pronged evaluation: automatic translation evaluation and user study.}
\\
\textit{\review{\aseupdate{Automatic Translation Evaluation:}}} 
To evaluate the performance of the automatic translation, we applied \ConfMigTool on "test-set" of 251 that we discussed in~\autoref{sub:sec:dataset}.
%
%
We evaluated two aspects of the automatic translation. 
First, we calculated the percentage of automated translations, which quantifies how many of the H-2 ASTs collected from each source CI file are matched and translated by \ConfMigTool.
Second, we adopted Cosine similarity~\cite{cosine_sim} and CrystalBLEU~\cite{Eghbali_2022} to measure the similarity between \ConfMigTool generated CI configuration files and developer-written Target CI configuration files.
We chose these two metrics due to their wide usage in literature. Cosine similarity is known for its versatility and applicability in source code migration research works~\cite{Xie_2020,Chen_Monperrus_2019, phan_2017, Talebipour_2021, Nguyen_2017}, and CrystalBLEU is designed for source code similarity and was utilized in code generation works~\cite{li2023codeeditor,yang2023syntax,dong2023codescore} and code migration works~\cite{pan2023understanding,jiao2023evaluation}.
For comparative analysis, we compared the performance of \ConfMigTool with that of \FirstPartyTool, the official tool from GitHub Actions~\cite{ghImporter}, using these two metrics. 
%

\textit{\review{User Study:}} 
We performed a user study to evaluate the practicality of \ConfMigTool.
The study was done with five participants, out of 15 initially contacted. They had software development or research experience ranging from 3 to 7 years, CI experience including GHA ranging from 1 month to 1 year. Each participant was tasked with migrating five Travis CI projects to GHA manually. They also migrated these projects semi-automatically twice, with one migration using a configuration file generated by \ConfMigTool and another with \FirstPartyTool.
The five projects are from the Travis CI-only set, and we selected them using popularity ($\ge 5$ stars or $\ge 5$ forks), project activity ( $ > 200$ commits made), and project freshness ( updated June 2023 or later)  following criteria in a similar 
works~\cite{Gousios_2017,Kalliamvakou_2016,Munaiah_2017}.
The five projects are \texttt{hutool}~\cite{dromara},  \texttt{WxJava}~\cite{Wechat-Group}, \texttt{hsweb-framework}~\cite{hsweb-framework_2024}, \texttt{elasticsearch-sql}~\cite{elasticsearch-sql_2024}, \texttt{TelegramBots}~\cite{Bermudez_2024}. 
For the study, we only considered Travis CI to GHA migrations as \FirstPartyTool only supports Travis CI to GHA.
All configuration files generated by \ConfMigTool and \FirstPartyTool were anonymized before being shared with the participants to avoid bias.
For each migration task, the participants were asked to achieve a "First Passing Workflow", a workflow that implements minimal CI functionality, and a "Final Workflow" which implements all CI functionality that is available in Source CI configuration.
During the study, we measured how much time can be saved via the semi-automatic migration approaches using \ConfMigTool and \FirstPartyTool generated files.
Also, we received ratings for the usefulness of the generated files by each tool from participants using a Likert scale~\cite{likert} ranging from 1 to 5, with 1 being "not at all useful" and 5 being "incredibly useful". 
Our full study guide and the full reports are available at
~\cite{repl}.

\subsection{RQ2: What is the \ConfMigTool Execution Cost?}
\label{sub:sec:approach-cost}
To \aseupdate{estimate} the time consumption of the training and translation processes, we programmatically measured the time it took to execute each training task, as well as the execution times for each translation performed on our test set, \aseupdate{during the experiment execution}. 
\aseupdate{We performed our experiment} on a machine running Ubuntu 22.04 and configured with an Intel Xeon CPU with 6 cores/12 threads and 32 GB of RAM. 

\subsection{RQ3: What are the Shortcomings of \ConfMigTool?}
\label{sub:sec:approach-issues}

\update{
To find the shortcomings of the results of \ConfMigTool, we asked co-authors who did not work on developing  \ConfMigTool to \aseupdate{manually evaluate the results of our experiments by} providing detailed reports on the different issues they noticed for the worst 25 generated translations from Travis CI to GHA, and the worst 25 generated translations from GHA to Travis CI, as determined by Cosine similarity. We then grouped similar issues into 3 main categories and reported the number of translations of each type that possessed that flaw.}
\section{Results}
\label{sec:results}

\subsection{\textbf{RQ1:} \ConfMigTool Migration \review{Effectiveness}}
\label{sub:sec:rq1}

\subsubsection{\review{Automatic Translation Evaluation Results}}

\begin{figure}[htbp]
\centering
  \includegraphics[width=0.87\linewidth,height=3cm]{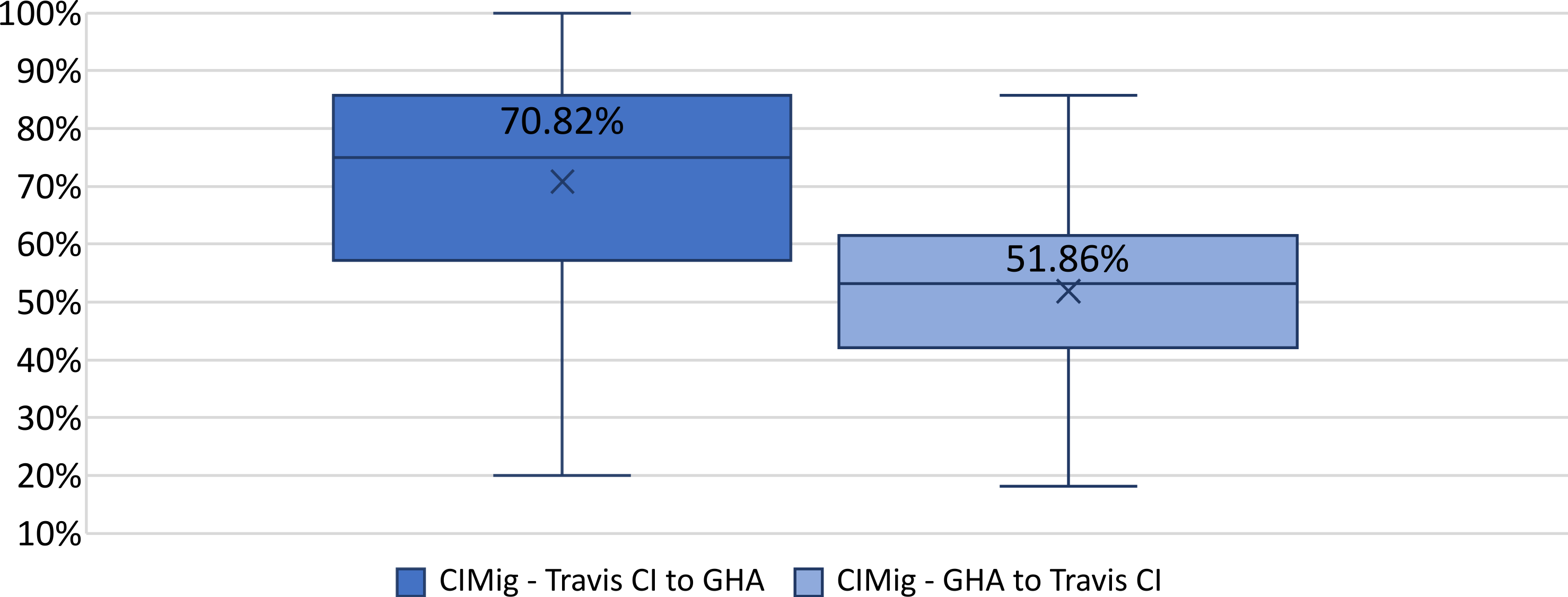}
\caption{Percentage of H-2 ASTs translated per-file}
    \label{fig:rq2:TransPerc}
\end{figure}

As the translation percentage values illustrate within~\autoref{fig:rq2:TransPerc}, \review{for the 251 projects composing the test set of \equivFileTuples,} our technique is effective at translating an average of \update{70.82\%} and a median of \update{75\%} of the H-2 ASTs extracted from a Travis CI file. \update{Furthermore, \ConfMigTool translates an average of 51.86\% and a median of 53.13\% of a GHA H-2 ASTs to Travis CI syntax.}

\begin{figure}[htbp]
    \scriptsize
    \centering
    \includegraphics[width=0.8\linewidth]{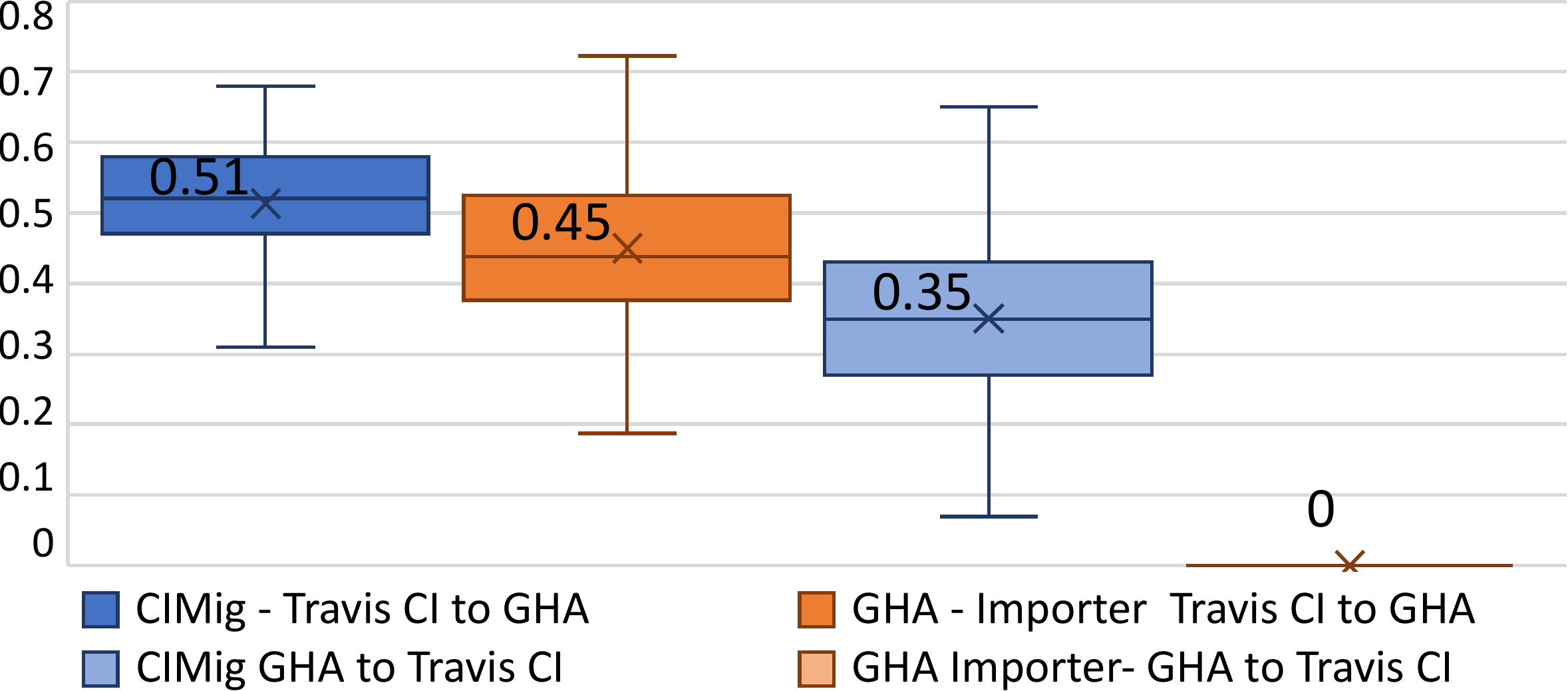}
    \caption{Cosine Similarity score of the Generated files.}
    \label{fig:rq2:Cosine}
\end{figure}

\begin{figure}[htbp]
\includegraphics[width=0.8\linewidth]{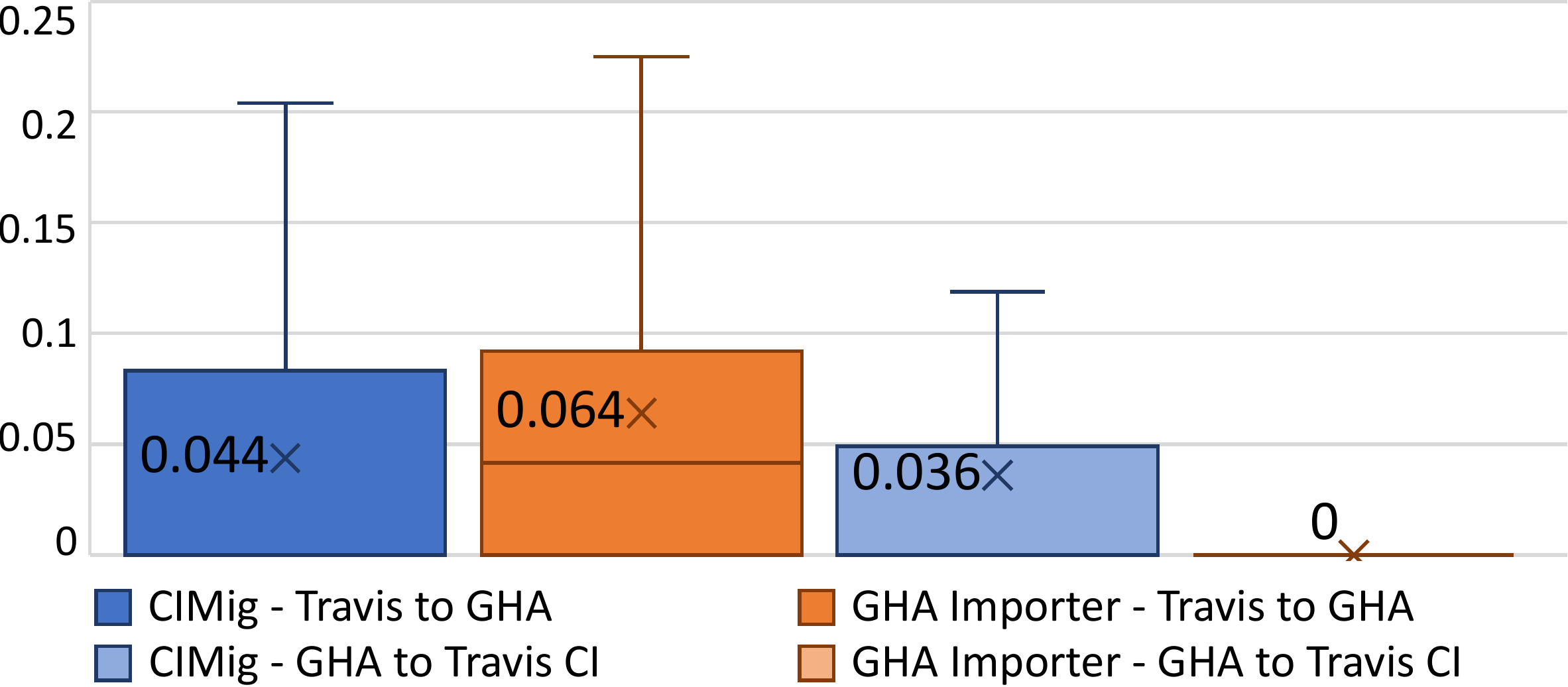}
\caption{CrystalBLEU score of the Generated files.}
\label{fig:rq2:BLEU}
\end{figure}

Moving on to the translation quality, both Cosine Similarity and CrystalBLEU scores, illustrated within~\autoref{fig:rq2:Cosine} and~\autoref{fig:rq2:BLEU} respectively,  \review{and measured using the test set,} display averages of 0.51 and 0.044 for the translation from Travis CI to GHA. The translation from GHA to Travis CI has averages of 0.35 and 0.036, respectively.

\update{Using the CrystalBLEU metric,} it's clear that the generated \NewGHAFiles and the \NewTrFiles show a good similarity to their developer-crafted baselines, especially when considering the works of Eghbali \& Pradel.~\cite{Eghbali_2022} \update{For CrystalBLEU}, where values around 0.05 were considered indicative of \update{context-preservation and feature-parity between the code pairs.}
\review{Furthermore}, Using both Cosine Similarity and CrystalBLEU to compare the results from \ConfMigTool to those obtained by \FirstPartyTool for the scenario of translating Travis CI files to GHA files, it's clear that \ConfMigTool's generated files are \review{ as similar to}
the developer-provided files. This helps substantiate the quality of our generated files, especially since the official tool relies on hand-crafted specific rules. While \ConfMigTool supports the translation from GHA to Travis CI, \FirstPartyTool does not, nor does any current tool, hence why there is no baseline \review{for comparison in that case.}

\update{
We find it important to mention that a considerable amount of \aseupdate{GHA} syntax does not have \aseupdate{a Travis CI equivalent}, mainly because the oft-used \texttt{Actions} in GHA do not have a direct equivalent in Travis CI, \update{as it doesn't support reusable workflows,} making their translation difficult, \update{explaining the lower results obtained when translating GHA files to Travis CI.}
} It's also notable that \equivFileTuples contains GHA and Travis CI tuples that have as little as 50\% functionality in common, meaning that the \NewGHAFile or \NewTrFile may only achieve a maximum similarity of 50\%. 


Overall, our results further support the confidence in the quality of the equivalent Target CI file generated by \ConfMigTool and validate that they implement a sizeable percentage of the functionality originally found in the Source CI file. We believe the files \ConfMigTool generates can form a good basis for developers to build on and help accelerate the migration process of their infrastructure, \review{and we attempt to confirm this in the following section.}



\renewcommand{\tabcolsep}{1.5pt}
\begin{table*}[htbp]
\caption{\review{User Study results on manual migration, and migrations with \ConfMigTool and with GHA Importer}}
\label{tab:man_eval_results}
\centering
\scriptsize
\begin{tabular}{c|rrrrr|rrrrr|rr}
\toprule
\multirow{4}{*}{ \makecell[c]{ Project name} }& \multicolumn{5}{c|}{First Workflow} & \multicolumn{5}{c|}{Final Workflow} & \multicolumn{2}{c}{Avg. user rating} \\
\cmidrule(r){2-6}
\cmidrule(r){7-11}
\cmidrule(r){12-13}
& Manual & \multicolumn{2}{c}{\ConfMigTool} & \multicolumn{2}{c|}{GHA Imp.}  & Manual & \multicolumn{2}{c}{\ConfMigTool} & \multicolumn{2}{c|}{GHA Imp.} & \multirow{3}{*}{ \makecell[c]{\ConfMigTool \\ files}} & \multirow{3}{*}{\makecell[r]{ GHA \\ Imp. \\files} }\\
\cmidrule(r){2-2}
\cmidrule(r){3-4}
\cmidrule(r){5-6}
\cmidrule(r){7-7}
\cmidrule(r){8-9}
\cmidrule(r){10-11}
& time (m) & time (m) & saved (\%) & time (m) & saved (\%) & time (m) & time (m) & saved (\%) & time (m) & saved (\%) & & \\

\midrule
WxJava& 38.40& 23.6 & 38.54 & 11.60 & 69.79 & 76.80 & 30.20 & 60.68 & 19.80 & 74.22 & 2.60   & 4.80 \\
hutool & 41.40 & 26.4 & 36.23 & 9.00 & 78.26 & 90.40 & 40.20 & 55.53 & 14.80 & 83.63 & 2.80 & 4.40 \\
Elasticsearch-sql & 29.00 & 24.20 & 16.55 & 4.60 & 84.14 & 46.00 & 36.20 & 21.30 & 8.00 & 82.61 & 3.40 & 5.00 \\
Hsweb-framework & 68.80 & 9.80 & 85.7 & 24.00 & 65.12 & 93.40 & 34.60 & 62.96 & 45.00 & 51.82 & 3.40 & 3.40 \\
Telegram Bots & 22.20 & 10.40 & 53.15 & 17.20 & 22.52 & 73.40 & 26.80 & 63.49 & 33.60 & 54.22 & 3.00 & 3.20 \\
\midrule
Average & 39.96 & 18.88 & 46.05\% & 13.28 & 63.97\%& 76 & 33.6& 52.79\%& 24.24& 69.30\%  & 3.04 & 4.16 \\
\bottomrule

\end{tabular}
\end{table*}

\subsubsection{\review{User Study}}
\review{
\autoref{tab:man_eval_results} shows the results of the user study conducted following Section \ref{sub:sec:accuracy_eval}. 
Column 1 shows the name of projects used for the migration from Travis CI to GHA in the study.
The First Workflow (column 2-6) shows the time that developers spent on Manual migration and migrations using \ConfMigTool and \FirstPartyTool to reach a First passing workflow.
For the results of \ConfMigTool and \FirstPartyTool, we show how much time was saved in comparison to Manual migration (column 4 and column 6).
Similarly, the Final Workflow (column 7-11) shows measures of developer time taken to reach a Final passing workflow for the same 3 migration types.
Finally, Avg. User Rating (column 12-13) contains the average usefulness scores from 1 to 5, assigned by the developers to the files from \ConfMigTool and \FirstPartyTool.}

\review{
The results show that using the \ConfMigTool or \FirstPartyTool helps developers reach both the First-passing workflow and the Final-passing workflow much faster than manual migrations. 
Indeed, \ConfMigTool reduced the Manual migration time by 16\% to 86\%, and \FirstPartyTool reduced it by 22\% to 84\% for reaching the First-passing workflow.
We see similar reductions as well in the migration time for the Final-passing workflow.
%
%
In terms of user ratings, \ConfMigTool has an average of 3.04 user rating, and \FirstPartyTool has a higher average user rating of 4.16. }
\review{
We notice similar user ratings for the two tools for the \texttt{Hsweb-framework} and \texttt{Telegram Bots} projects. In both projects, \ConfMigTool also provides higher reduction in migration time than \FirstPartyTool. Two developers mentioned in their reports that the files provided by \ConfMigTool were easier to extend for these 2 projects than \FirstPartyTool's files, where \FirstPartyTool's attempts to translate some syntax results in more complicated configuration files.
%
%
%
In summary, \ConfMigTool shows lower reduction rate than \FirstPartyTool on three projects and higher reduction rate on two projects. 
User ratings tend to follow the reduction rates, with ratings for \ConfMigTool being lower than \FirstPartyTool.
Overall, the results confirm that the files generated by \ConfMigTool are usable in the GHA environment with minor modifications, and help save on migration time.
%
%
Furthermore, \ConfMigTool leverages mining processes from existing files,
making it easy to extend and adapt to new syntax,
as shown via GHA to Travis CI migration. 
On the other hand, \FirstPartyTool is built using manual-mapped rules,
%
and only supports Travis CI to GHA migration, limiting its extension to other scenarios.
Hence, \ConfMigTool provides more usefulness in terms of supporting more migration scenarios with comparable performance to the specialized migration tool.
}

\subsection{\textbf{RQ2:} \ConfMigTool Execution Cost}
\label{sub:sec:rq3}
Concerning the time needed to perform the translation of GHA files to Travis CI the average execution time of \ConfMigTool is 719.85 milliseconds, and the median is 705 milliseconds. Meanwhile, the average execution time of \FirstPartyTool is 1553.31 milliseconds, and the median is 1503.38, \aseupdate{for} the same process. \aseupdate{While both} executions times are acceptable~\cite{Response_time}, \ConfMigTool is faster than \review{\FirstPartyTool}. \aseupdate{\ConfMigTool has acceptable times for} translating GHA syntax to Travis CI as well, with a median execution time of 797 milliseconds, and an average translation time of 1235.46 milliseconds. 

Concerning the different processes of the training phase, they are only executed once and are independent of the translation process, making their time consumption less important. During the rule mining phase, detailed in  ~\autoref{sub:sec:rule_mining}, we executed 2 Apriori-based ARM operations: Travis CI to GHA translation rules, which took 45947 milliseconds to execute, and GHA  hierarchization rules which took 22022 milliseconds to execute. \update{These times were nearly identical when mining the translation rules GHA to Travis CI, as well as the generation of Travis CI hierarchization rules.} The most time-consuming process we designed was the detection of which GHA and Travis CI Frequent Trees match with the Stat-Based rules, which took 1625773 milliseconds (around 27 minutes), despite a parallelized implementation that took advantage of all CPU threads available. This is however not surprising as there is a total of \totalNonSimBasedRules Stat-Based rules for each direction, along with \totalTARGeneratedTravis Travis CI Frequent Trees and \totalTARGeneratedGHA GHA Frequent Trees. 

We performed two Frequent-Trees mining operations: one on Travis CI files, which took 1211342 milliseconds (around 20 minutes), and one on GitHub Actions files which took 257445 seconds (around 71 hours). The latter's much higher time consumption can be attributed to a bigger number of unique root nodes at which we attempted to detect Frequent-Trees, as well as the larger and more complex ASTs of GHA files. Overall, we believe \ConfMigTool time consumption during the training phase also remains within acceptable limits. 

\subsection{\textbf{RQ3:} \ConfMigTool Translation Failures}
\label{sub:sec:rq4}
Although our approach generates CI files of good quality, there are certain cases where our approach fails to generate an acceptable translation. These failures \review{are} classified into three categories \review{as follows}:

\noindent\textbf{Syntax with no direct equivalent:} (5 out of 25 Travis CI $\Rightarrow$ GHA translations, 22 out of 25 GHA $\Rightarrow$ Travis CI  translations) Although there are some similarities between Travis  CI and GHA configuration syntaxes, there are certain functionalities that are supported in only one of them.
For example, 
GHA offers the \texttt{uses} keyword that allows reuse of existing GHA workflows in the form of Actions, but Travis CI does not offer an equivalent functionality.
\review{An example of this syntax is shown in~\hyperref[lst:gh_to_tr]{Listing 1}.} 

\begin{minted}[fontsize=\scriptsize]{yaml}
uses: sonarsource/sonarcloud-github-action
env:
    GITHUB_TOKEN: ${{ secrets.GITHUB_TOKEN }}
    SONAR_TOKEN: ${{ secrets.SONAR_TOKEN }}
    SONAR_SCANNER_OPTS: -Dsonar.organization=albertus82-github
\end{minted}
\captionof{listing}{\review{GHA syntax with no Travis CI equivalent from \texttt{Albertus82/Cyclesmod}}}
\label{lst:gh_to_tr}

\noindent\textbf{Syntax that relies on more than two levels:} (7 out of 25 Travis CI $\Rightarrow$ GHA translations, 2 out of 25 GHA $\Rightarrow$ Travis CI translations) Since we opted to capture and translate H-2 ASTs in Travis CI, any functionalities that depend on the configuration of more than 2 levels are not captured. For example, the usage of multiple stages with different \texttt{jdk} and \texttt{language} settings in Travis.\\

\noindent\textbf{Unabstracted syntax and parsing issues:} (23 out of 25 Travis CI $\Rightarrow$ GHA translations, 4 out of 25 GHA $\Rightarrow$ Travis CI translations) Since the abstraction process was applied with the usage of the most common commands, some less common commands, such as \texttt{openssl} and \texttt{jarsigner},
are not represented within the translation rules we generated. \review{An illustration of this is shown in~\hyperref[lst:tr_to_gh]{Listing 2}}.

\begin{minted}[fontsize=\scriptsize,escapeinside=||]{yaml}
before_deploy:
    |\colorbox{yellow}{openssl}| aes-256-cbc -K $encrypted_key 
         -iv $encrypted_iv -in release-key.jks.enc 
         -out gotify-release-key.jks -d  ...
    |\colorbox{yellow}{jarsigner}| -verbose -sigalg SHA1withRSA 
        -digestalg SHA1 -keystore release-key.jks
\end{minted}
\captionof{listing}{\review{Travis CI syntax from \texttt{Gotify/Android} containing \colorbox{yellow}{unmatched commands}}}
\label{lst:tr_to_gh}


\section{Related works}
\label{sec:related_works}
\noindent\textbf{Automatic Code Migration.}
Migrating from one programming language to another is very common in large software systems due to the need for cross-platform support and language support features. However, programming language migration is effort-intensive and error-prone~\cite{Zhong2010,Meng2012,chen2020similarapi} due to the differences in syntax and unfamiliarity with the target programming language. To mitigate this, researchers developed tools and techniques for automatic programming language migration. For example, Java2CSharp~\cite{Java2CSharp} and j2swift~\cite{j2swift} are developed for migrating from Java to C\# and Swift. However, these tools and other research works~\cite{Ramly2006, Mossienko2003, hassan2005lightweight} use predefined transformation rules for their migration. Creating these rules is a laborious process, and in many cases, these migrations may fail due to complex and rare syntax used by different programming languages. To resolve these limitations, Zhong et al.~\cite{Zhong2010} and Nguyen et al.~\cite{nguyen2014statistical} utilized a mining-based approach for automatic migration. These approaches heavily relied on similarity-based alignment and may not correctly migrate code if the target language adopts a different naming scheme. 
mppSMT by Nguyen et al.~\cite{Nguyen2016}, utilizes a divide-and-conquer approach with a phrase-based SMT engine to integrate the semantic features for automatic migration. The approach uses data and control dependency of source code, which may not be applicable to configuration code due to its higher level of abstraction. j2sInferer~\cite{An2018} is a recent approach that utilizes syntax and mapping rules with minimal domain logic for migration of Android Java code to Swift code with 65\% cross-project accuracy. Such syntax similarity is very low among configuration code files and makes alignment infeasible. More recently, ML-based techniques~\cite{chen2018tree,gu2017deepam} are proposed for the automatic migration of programming languages. However, ML approaches require large corpora for model training, which may not be feasible for recently developed programming languages or DevOps configuration files where very little migration data exists. 

\noindent\textbf{Configuration Maintenance.} Like source code files, the different configuration code files for CI systems, Build systems, etc., are integral parts of software projects. Prior works suggested that developers often work on maintaining and migrating configuration systems~\cite{widder2018ci,zhang2018cd, Rzig_2022, gligoric2014automated} to improve performance and productivity. However, maintaining configuration code is tedious due to limited domain-specific knowledge and syntactical differences in configuration code across different tools. Gligoric et al.~\cite{gligoric2014automated} utilized dynamic analysis and search-based refactoring techniques to automatically migrate build systems. Moreover, automated program repair-based techniques~\cite{hassan2018hirebuild, lou2019history, mukherjee2021fixing} are applied to fix build scripts. Xue et al.~\cite{xu2020sledge} proposed a technique for automatic migration to Docker containers.
Recently, Henkel et al.~\cite{Henkel_2020} proposed binnacle to automatically detect bad practices in Docker files.
Vassallo et al.~\cite{vassallo2019automated} utilized program analysis techniques to detect anti-patterns in CI configuration scripts. 
At the same time, Rahman \& Parmin~\cite{Rahman2023} proposed a technique for automatically detecting security vulnerabilities in Puppet-based IaC configurations. Although there are several techniques for the automatic migration and maintenance of different configuration systems, there is no research work on the automatic migration of CI systems.  

\section{Threats to Validity}
\label{sec:threats}

\noindent\textbf{Internal Validity.} The main threat is the incorrect composition of the generated CI configuration code. To mitigate this, we tested our approach thoroughly in several rounds, and we contextualized our results by comparing them to both developer-crafted and \FirstPartyTool-generated files.
We also evaluated the generated files with state-of-the-art metrics to evaluate the correctness of the approach\review{, and further evaluated them via the user study.}

\noindent\textbf{External Validity.} We evaluated our approach for migration between Travis CI and GitHub Actions. These projects are Java-based and OSS in nature. So, our approach may not work correctly on other CI systems with different programming languages and closed-source projects. Although the evaluation is CI system-specific, the proposed rule mining and composition techniques are more generic. Moreover, different CI systems support similar functionalities and similar structures\review{, such as YAML}. So, we believe that our proposed approach will work for other CI systems as well\review{, with sufficient retraining}. \review{ We attempted to approximate actual user experience via our user study by recruiting developers with varied development and CI experiences, but their experiences may not reflect every possible users'.}

\noindent\textbf{Construction Validity.} For automatic rule generation, we considered two-level (H-2) level AST transition nodes.
\review{We believe these} rules are a good balance between conservativeness and diversity\review{, for reasons detailed in the Parameter Tuning paragraph of ~\ref{par:parameter_tuning}}.
\section{Conclusion and Future work }
\label{sec:conclusion}
With the growing use of CI systems for faster code integration, migration of CI systems has become very common in development activity. However, migrating CI systems is a tedious and error-prone process~\cite{Mazrae_2023}. We presented \ConfMigTool To assist the developers with CI migration,  and help facilitate this process. In our evaluation, even with a small set of existing CI migration data, \ConfMigTool can generate \aseupdate{CI files}
of good similarity to the developer-crafted versions. \review {Furthermore, the user study also suggests that \ConfMigTool is beneficial for developers, allowing them to migrate CI systems in less time than manual migration.} \review{Moreover}, the proposed approach is generic in nature and can be easily applied to other configuration systems as well. In
the future, we plan to incorporate large language models (LLM), such as ChatGPT, to generate more accurate migration rules and apply the automatic migration process to other configuration systems, such as Docker, etc.

\newpage
\bibliography{Bibliography.bib}


\begin{thebibliography}{87}


\ifx \showCODEN    \undefined \def \showCODEN     #1{\unskip}     \fi
\ifx \showDOI      \undefined \def \showDOI       #1{#1}\fi
\ifx \showISBNx    \undefined \def \showISBNx     #1{\unskip}     \fi
\ifx \showISBNxiii \undefined \def \showISBNxiii  #1{\unskip}     \fi
\ifx \showISSN     \undefined \def \showISSN      #1{\unskip}     \fi
\ifx \showLCCN     \undefined \def \showLCCN      #1{\unskip}     \fi
\ifx \shownote     \undefined \def \shownote      #1{#1}          \fi
\ifx \showarticletitle \undefined \def \showarticletitle #1{#1}   \fi
\ifx \showURL      \undefined \def \showURL       {\relax}        \fi
\providecommand\bibfield[2]{#2}
\providecommand\bibinfo[2]{#2}
\providecommand\natexlab[1]{#1}
\providecommand\showeprint[2][]{arXiv:#2}

\bibitem[Agrawal et~al\mbox{.}(1994)]%
        {agrawal1994fast}
\bibfield{author}{\bibinfo{person}{Rakesh Agrawal}, \bibinfo{person}{Ramakrishnan Srikant}, {et~al\mbox{.}}} \bibinfo{year}{1994}\natexlab{}.
\newblock \showarticletitle{Fast algorithms for mining association rules}. In \bibinfo{booktitle}{\emph{Proc. 20th int. conf. very large data bases, VLDB}}, Vol.~\bibinfo{volume}{1215}. \bibinfo{pages}{487--499}.
\newblock


\bibitem[Ahmad et~al\mbox{.}(2021)]%
        {ahmad-2021}
\bibfield{author}{\bibinfo{person}{Wasi Ahmad}, \bibinfo{person}{Saikat Chakraborty}, \bibinfo{person}{Baishakhi Ray}, {and} \bibinfo{person}{Kai-Wei Chang}.} \bibinfo{year}{2021}\natexlab{}.
\newblock \showarticletitle{Unified Pre-training for Program Understanding and Generation}. In \bibinfo{booktitle}{\emph{Proceedings of the 2021 Conference of the North American Chapter of the Association for Computational Linguistics: Human Language Technologies}}. \bibinfo{publisher}{Association for Computational Linguistics}, \bibinfo{address}{Online}, \bibinfo{pages}{2655--2668}.
\newblock
\urldef\tempurl%
\url{https://www.aclweb.org/anthology/2021.naacl-main.211}
\showURL{%
\tempurl}


\bibitem[Albaum(1997)]%
        {likert}
\bibfield{author}{\bibinfo{person}{Gerald Albaum}.} \bibinfo{year}{1997}\natexlab{}.
\newblock \showarticletitle{The Likert Scale Revisited}.
\newblock \bibinfo{journal}{\emph{Market Research Society. Journal.}} \bibinfo{volume}{39}, \bibinfo{number}{2} (\bibinfo{year}{1997}), \bibinfo{pages}{1--21}.
\newblock
\urldef\tempurl%
\url{https://doi.org/10.1177/147078539703900202}
\showDOI{\tempurl}
\showeprint{https://doi.org/10.1177/147078539703900202}


\bibitem[An et~al\mbox{.}(2018)]%
        {An2018}
\bibfield{author}{\bibinfo{person}{Kijin An}, \bibinfo{person}{Na Meng}, {and} \bibinfo{person}{Eli Tilevich}.} \bibinfo{year}{2018}\natexlab{}.
\newblock \showarticletitle{Automatic Inference of Java-to-Swift Translation Rules for Porting Mobile Applications}. In \bibinfo{booktitle}{\emph{2018 IEEE/ACM 5th International Conference on Mobile Software Engineering and Systems (MOBILESoft)}}. \bibinfo{pages}{180--190}.
\newblock


\bibitem[Anonymous(2024)]%
        {repl}
\bibfield{author}{\bibinfo{person}{Anonymous}.} \bibinfo{year}{2024}\natexlab{}.
\newblock \bibinfo{title}{Replication Package}.
\newblock
\newblock
\urldef\tempurl%
\url{https://figshare.com/s/d903576fab38e2a54660}
\showURL{%
\tempurl}


\bibitem[Azure(2023a)]%
        {azure}
\bibfield{author}{\bibinfo{person}{Azure}.} \bibinfo{year}{2023}\natexlab{a}.
\newblock \bibinfo{title}{Azure Pipelines | Microsoft Azure}.
\newblock
\newblock
\urldef\tempurl%
\url{https://azure.microsoft.com/en-us/products/devops/pipelines}
\showURL{%
\tempurl}


\bibitem[Azure(2023b)]%
        {juliakm_2023}
\bibfield{author}{\bibinfo{person}{Microsoft Azure}.} \bibinfo{year}{2023}\natexlab{b}.
\newblock \bibinfo{title}{Build and Release Tasks - Azure Pipelines}.
\newblock
\newblock
\urldef\tempurl%
\url{https://learn.microsoft.com/en-us/azure/devops/pipelines/process/tasks}
\showURL{%
\tempurl}


\bibitem[Azure(2023c)]%
        {steved0x_2023}
\bibfield{author}{\bibinfo{person}{Microsoft Azure}.} \bibinfo{year}{2023}\natexlab{c}.
\newblock \bibinfo{title}{Microsoft-hosted agents for Azure Pipelines - Azure Pipelines}.
\newblock
\newblock
\urldef\tempurl%
\url{https://learn.microsoft.com/en-us/azure/devops/pipelines/agents/hosted}
\showURL{%
\tempurl}


\bibitem[Beller et~al\mbox{.}(2017)]%
        {Beller_2017}
\bibfield{author}{\bibinfo{person}{Moritz Beller}, \bibinfo{person}{Georgios Gousios}, {and} \bibinfo{person}{Andy Zaidman}.} \bibinfo{year}{2017}\natexlab{}.
\newblock \showarticletitle{Oops, My Tests Broke the Build: An Explorative Analysis of Travis CI with GitHub}. In \bibinfo{booktitle}{\emph{2017 IEEE/ACM 14th International Conference on Mining Software Repositories (MSR)}}. \bibinfo{publisher}{IEEE}, \bibinfo{address}{Buenos Aires, Argentina}, \bibinfo{pages}{356–367}.
\newblock
\showISBNx{978-1-5386-1544-7}
\urldef\tempurl%
\url{https://doi.org/10.1109/MSR.2017.62}
\showDOI{\tempurl}


\bibitem[Bermudez(2024)]%
        {Bermudez_2024}
\bibfield{author}{\bibinfo{person}{Ruben Bermudez}.} \bibinfo{year}{2024}\natexlab{}.
\newblock \bibinfo{title}{rubenlagus/TelegramBots}.
\newblock
\newblock
\urldef\tempurl%
\url{https://github.com/rubenlagus/TelegramBots}
\showURL{%
\tempurl}


\bibitem[Bertram(2021)]%
        {octopus}
\bibfield{author}{\bibinfo{person}{Adam Bertram}.} \bibinfo{year}{2021}\natexlab{}.
\newblock \bibinfo{title}{Config as Code: What is it and how is it beneficial?}
\newblock
\newblock
\urldef\tempurl%
\url{https://octopus.com/blog/config-as-code-what-is-it-how-is-it-beneficial}
\showURL{%
\tempurl}


\bibitem[Bird et~al\mbox{.}(2009)]%
        {Bird_2009}
\bibfield{author}{\bibinfo{person}{Christian Bird}, \bibinfo{person}{Peter~C. Rigby}, \bibinfo{person}{Earl~T. Barr}, \bibinfo{person}{David~J. Hamilton}, \bibinfo{person}{Daniel~M. German}, {and} \bibinfo{person}{Prem Devanbu}.} \bibinfo{year}{2009}\natexlab{}.
\newblock \showarticletitle{The promises and perils of mining git}. In \bibinfo{booktitle}{\emph{2009 6th IEEE International Working Conference on Mining Software Repositories}}. \bibinfo{publisher}{IEEE}, \bibinfo{address}{Vancouver, BC, Canada}, \bibinfo{pages}{1–10}.
\newblock
\showISBNx{978-1-4244-3493-0}
\urldef\tempurl%
\url{https://doi.org/10.1109/MSR.2009.5069475}
\showDOI{\tempurl}


\bibitem[Chen(2020)]%
        {chen2020similarapi}
\bibfield{author}{\bibinfo{person}{Chunyang Chen}.} \bibinfo{year}{2020}\natexlab{}.
\newblock \showarticletitle{Similarapi: mining analogical apis for library migration}. In \bibinfo{booktitle}{\emph{Proceedings of the ACM/IEEE 42nd International Conference on Software Engineering: Companion Proceedings}}. \bibinfo{pages}{37--40}.
\newblock


\bibitem[Chen et~al\mbox{.}(2018)]%
        {chen2018tree}
\bibfield{author}{\bibinfo{person}{Xinyun Chen}, \bibinfo{person}{Chang Liu}, {and} \bibinfo{person}{Dawn Song}.} \bibinfo{year}{2018}\natexlab{}.
\newblock \showarticletitle{Tree-to-tree neural networks for program translation}.
\newblock \bibinfo{journal}{\emph{Advances in neural information processing systems}}  \bibinfo{volume}{31} (\bibinfo{year}{2018}).
\newblock


\bibitem[Chen and Monperrus(2019)]%
        {Chen_Monperrus_2019}
\bibfield{author}{\bibinfo{person}{Zimin Chen} {and} \bibinfo{person}{Martin Monperrus}.} \bibinfo{year}{2019}\natexlab{}.
\newblock \showarticletitle{The Remarkable Role of Similarity in Redundancy-based Program Repair}.
\newblock  \bibinfo{number}{arXiv:1811.05703} (\bibinfo{date}{May} \bibinfo{year}{2019}).
\newblock
\urldef\tempurl%
\url{http://arxiv.org/abs/1811.05703}
\showURL{%
\tempurl}
\newblock
\shownote{arXiv:1811.05703 [cs]}.


\bibitem[Chi et~al\mbox{.}(2005a)]%
        {CMAlg}
\bibfield{author}{\bibinfo{person}{Yun Chi}, \bibinfo{person}{Yi Xia}, \bibinfo{person}{Yirong Yang}, {and} \bibinfo{person}{Richard~R. Muntz}.} \bibinfo{year}{2005}\natexlab{a}.
\newblock \showarticletitle{Mining Closed and Maximal Frequent Subtrees from Databases of Labeled Rooted Trees}.
\newblock \bibinfo{journal}{\emph{IEEE Trans. Knowl. Data Eng.}} \bibinfo{volume}{17}, \bibinfo{number}{2} (\bibinfo{year}{2005}), \bibinfo{pages}{190--202}.
\newblock
\urldef\tempurl%
\url{https://doi.org/10.1109/TKDE.2005.30}
\showDOI{\tempurl}


\bibitem[Chi et~al\mbox{.}(2005b)]%
        {Chi_2005}
\bibfield{author}{\bibinfo{person}{Yun Chi}, \bibinfo{person}{Yirong Yang}, {and} \bibinfo{person}{Richard~R. Muntz}.} \bibinfo{year}{2005}\natexlab{b}.
\newblock \showarticletitle{Canonical forms for labelled trees and their applications in frequent subtree mining}.
\newblock \bibinfo{journal}{\emph{Knowledge and Information Systems}} \bibinfo{volume}{8}, \bibinfo{number}{2} (\bibinfo{date}{Aug} \bibinfo{year}{2005}), \bibinfo{pages}{203–234}.
\newblock
\showISSN{0219-1377, 0219-3116}
\urldef\tempurl%
\url{https://doi.org/10.1007/s10115-004-0180-7}
\showDOI{\tempurl}


\bibitem[Circle-CI(2023a)]%
        {oscircleci}
\bibfield{author}{\bibinfo{person}{Circle-CI}.} \bibinfo{year}{2023}\natexlab{a}.
\newblock \bibinfo{title}{Introduction to YAML Configurations - CircleCI}.
\newblock
\newblock
\urldef\tempurl%
\url{https://circleci.com/docs/introduction-to-yaml-configurations/}
\showURL{%
\tempurl}


\bibitem[Circle-CI(2023b)]%
        {orbs}
\bibfield{author}{\bibinfo{person}{Circle-CI}.} \bibinfo{year}{2023}\natexlab{b}.
\newblock \bibinfo{title}{Orbs overview - CircleCI}.
\newblock
\newblock
\urldef\tempurl%
\url{https://circleci.com/docs/orb-intro/}
\showURL{%
\tempurl}


\bibitem[CircleCI(2023)]%
        {circleci}
\bibfield{author}{\bibinfo{person}{CircleCI}.} \bibinfo{year}{2023}\natexlab{}.
\newblock \bibinfo{title}{Continuous Integration and Delivery - CircleCI}.
\newblock
\newblock
\urldef\tempurl%
\url{https://circleci.com/}
\showURL{%
\tempurl}


\bibitem[Dong et~al\mbox{.}(2023)]%
        {dong2023codescore}
\bibfield{author}{\bibinfo{person}{Yihong Dong}, \bibinfo{person}{Jiazheng Ding}, \bibinfo{person}{Xue Jiang}, \bibinfo{person}{Ge Li}, \bibinfo{person}{Zhuo Li}, {and} \bibinfo{person}{Zhi Jin}.} \bibinfo{year}{2023}\natexlab{}.
\newblock \showarticletitle{Codescore: Evaluating code generation by learning code execution}.
\newblock \bibinfo{journal}{\emph{arXiv preprint arXiv:2301.09043}} (\bibinfo{year}{2023}).
\newblock


\bibitem[dromara(2024)]%
        {dromara}
\bibfield{author}{\bibinfo{person}{dromara}.} \bibinfo{year}{2024}\natexlab{}.
\newblock
\newblock
\urldef\tempurl%
\url{https://github.com/dromara/hutool}
\showURL{%
\tempurl}


\bibitem[Durieux et~al\mbox{.}(2019)]%
        {Durieux_2019}
\bibfield{author}{\bibinfo{person}{Thomas Durieux}, \bibinfo{person}{Rui Abreu}, \bibinfo{person}{Martin Monperrus}, \bibinfo{person}{Tegawendé~F. Bissyandé}, {and} \bibinfo{person}{Luís Cruz}.} \bibinfo{year}{2019}\natexlab{}.
\newblock \showarticletitle{An Analysis of 35+ Million Jobs of Travis CI}.
\newblock \bibinfo{journal}{\emph{2019 IEEE International Conference on Software Maintenance and Evolution (ICSME)}} (\bibinfo{date}{Sep} \bibinfo{year}{2019}), \bibinfo{pages}{291–295}.
\newblock
\urldef\tempurl%
\url{https://doi.org/10.1109/icsme.2019.00044}
\showDOI{\tempurl}
\newblock
\shownote{arXiv: 1904.09416}.


\bibitem[Eghbali and Pradel(2022)]%
        {Eghbali_2022}
\bibfield{author}{\bibinfo{person}{Aryaz Eghbali} {and} \bibinfo{person}{Michael Pradel}.} \bibinfo{year}{2022}\natexlab{}.
\newblock \showarticletitle{CrystalBLEU: Precisely and Efficiently Measuring the Similarity of Code}. In \bibinfo{booktitle}{\emph{Proceedings of the 37th IEEE/ACM International Conference on Automated Software Engineering}}. \bibinfo{publisher}{ACM}, \bibinfo{address}{Rochester MI USA}, \bibinfo{pages}{1--12}.
\newblock
\showISBNx{978-1-4503-9475-8}
\urldef\tempurl%
\url{https://doi.org/10.1145/3551349.3556903}
\showDOI{\tempurl}


\bibitem[El-Ramly et~al\mbox{.}(2006)]%
        {Ramly2006}
\bibfield{author}{\bibinfo{person}{M. El-Ramly}, \bibinfo{person}{R. Eltayeb}, {and} \bibinfo{person}{H.A. Alla}.} \bibinfo{year}{2006}\natexlab{}.
\newblock \showarticletitle{An Experiment in Automatic Conversion of Legacy Java Programs to C\#}. In \bibinfo{booktitle}{\emph{IEEE International Conference on Computer Systems and Applications, 2006.}} \bibinfo{pages}{1037--1045}.
\newblock
\urldef\tempurl%
\url{https://doi.org/10.1109/AICCSA.2006.205215}
\showDOI{\tempurl}


\bibitem[Etemadi et~al\mbox{.}(2017)]%
        {etemadi2017association}
\bibfield{author}{\bibinfo{person}{Vahid Etemadi}, \bibinfo{person}{Omid Bushehrian}, {and} \bibinfo{person}{Reza Akbari}.} \bibinfo{year}{2017}\natexlab{}.
\newblock \showarticletitle{Association rule mining for finding usability problem patterns: A case study on StackOverflow}. In \bibinfo{booktitle}{\emph{2017 International Symposium on Computer Science and Software Engineering Conference (CSSE)}}. IEEE, \bibinfo{pages}{24--29}.
\newblock


\bibitem[Fau~Alexandre(2023)]%
        {Java2CSharp}
\bibfield{author}{\bibinfo{person}{Mauceri~Christian Fau~Alexandre}.} \bibinfo{year}{2023}\natexlab{}.
\newblock \bibinfo{title}{Java2CSharp}.
\newblock
\newblock
\urldef\tempurl%
\url{http://sourceforge.net/projects/j2cstranslator/}
\showURL{%
\tempurl}
\newblock
\shownote{accessed 04-01-2023}.


\bibitem[GitHub(2023a)]%
        {GitHubActionImporter}
\bibfield{author}{\bibinfo{person}{GitHub}.} \bibinfo{year}{2023}\natexlab{a}.
\newblock
\newblock
\urldef\tempurl%
\url{https://github.com/github/gh-actions-importer}
\showURL{%
\tempurl}


\bibitem[GitHub(2023b)]%
        {runs}
\bibfield{author}{\bibinfo{person}{GitHub}.} \bibinfo{year}{2023}\natexlab{b}.
\newblock
\newblock
\urldef\tempurl%
\url{https://docs.github.com/en/actions/using-workflows/workflow-syntax-for-github-actions#jobsjob_idruns-on}
\showURL{%
\tempurl}


\bibitem[GitHub(2023c)]%
        {uses}
\bibfield{author}{\bibinfo{person}{GitHub}.} \bibinfo{year}{2023}\natexlab{c}.
\newblock
\newblock
\urldef\tempurl%
\url{https://docs.github.com/en/actions/using-workflows/reusing-workflows}
\showURL{%
\tempurl}


\bibitem[GitHub(2023d)]%
        {github_website}
\bibfield{author}{\bibinfo{person}{GitHub}.} \bibinfo{year}{2023}\natexlab{d}.
\newblock \bibinfo{title}{About GitHub}.
\newblock
\newblock
\urldef\tempurl%
\url{https://github.com/about}
\showURL{%
\tempurl}


\bibitem[GitHub(2023e)]%
        {gha}
\bibfield{author}{\bibinfo{person}{GitHub}.} \bibinfo{year}{2023}\natexlab{e}.
\newblock \bibinfo{title}{GitHub Actions}.
\newblock
\newblock
\urldef\tempurl%
\url{https://github.com/features/actions}
\showURL{%
\tempurl}


\bibitem[GitHub(2023f)]%
        {ghImporter}
\bibfield{author}{\bibinfo{person}{GitHub}.} \bibinfo{year}{2023}\natexlab{f}.
\newblock \bibinfo{title}{github/gh-actions-importer}.
\newblock
\newblock
\urldef\tempurl%
\url{https://github.com/github/gh-actions-importer}
\showURL{%
\tempurl}


\bibitem[Gligoric et~al\mbox{.}(2014)]%
        {gligoric2014automated}
\bibfield{author}{\bibinfo{person}{Milos Gligoric}, \bibinfo{person}{Wolfram Schulte}, \bibinfo{person}{Chandra Prasad}, \bibinfo{person}{Danny Van~Velzen}, \bibinfo{person}{Iman Narasamdya}, {and} \bibinfo{person}{Benjamin Livshits}.} \bibinfo{year}{2014}\natexlab{}.
\newblock \showarticletitle{Automated migration of build scripts using dynamic analysis and search-based refactoring}.
\newblock \bibinfo{journal}{\emph{ACM SIGPLAN Notices}} \bibinfo{volume}{49}, \bibinfo{number}{10} (\bibinfo{year}{2014}), \bibinfo{pages}{599--616}.
\newblock


\bibitem[Golzadeh et~al\mbox{.}(2022)]%
        {Golzadeh_2022}
\bibfield{author}{\bibinfo{person}{Mehdi Golzadeh}, \bibinfo{person}{Alexandre Decan}, {and} \bibinfo{person}{Tom Mens}.} \bibinfo{year}{2022}\natexlab{}.
\newblock \showarticletitle{On the rise and fall of CI services in GitHub}. In \bibinfo{booktitle}{\emph{2022 IEEE International Conference on Software Analysis, Evolution and Reengineering (SANER)}}. \bibinfo{publisher}{IEEE}, \bibinfo{address}{Honolulu, HI, USA}, \bibinfo{pages}{662–672}.
\newblock
\showISBNx{978-1-66543-786-8}
\urldef\tempurl%
\url{https://doi.org/10.1109/SANER53432.2022.00084}
\showDOI{\tempurl}


\bibitem[Gousios and Spinellis(2017)]%
        {Gousios_2017}
\bibfield{author}{\bibinfo{person}{Georgios Gousios} {and} \bibinfo{person}{Diomidis Spinellis}.} \bibinfo{year}{2017}\natexlab{}.
\newblock \showarticletitle{Mining Software Engineering Data from GitHub}. In \bibinfo{booktitle}{\emph{2017 IEEE/ACM 39th International Conference on Software Engineering Companion (ICSE-C)}}. \bibinfo{publisher}{IEEE}, \bibinfo{address}{Buenos Aires, Argentina}, \bibinfo{pages}{501–502}.
\newblock
\showISBNx{978-1-5386-1589-8}
\urldef\tempurl%
\url{https://doi.org/10.1109/ICSE-C.2017.164}
\showDOI{\tempurl}


\bibitem[Gu et~al\mbox{.}(2017)]%
        {gu2017deepam}
\bibfield{author}{\bibinfo{person}{Xiaodong Gu}, \bibinfo{person}{Hongyu Zhang}, \bibinfo{person}{Dongmei Zhang}, {and} \bibinfo{person}{Sunghun Kim}.} \bibinfo{year}{2017}\natexlab{}.
\newblock \showarticletitle{DeepAM: Migrate APIs with multi-modal sequence to sequence learning}.
\newblock \bibinfo{journal}{\emph{arXiv preprint arXiv:1704.07734}} (\bibinfo{year}{2017}).
\newblock


\bibitem[Hassan and Holt(2005)]%
        {hassan2005lightweight}
\bibfield{author}{\bibinfo{person}{Ahmed~E Hassan} {and} \bibinfo{person}{Richard~C Holt}.} \bibinfo{year}{2005}\natexlab{}.
\newblock \showarticletitle{A lightweight approach for migrating Web frameworks}.
\newblock \bibinfo{journal}{\emph{Information and Software Technology}} \bibinfo{volume}{47}, \bibinfo{number}{8} (\bibinfo{year}{2005}), \bibinfo{pages}{521--532}.
\newblock


\bibitem[Hassan and Wang(2018)]%
        {hassan2018hirebuild}
\bibfield{author}{\bibinfo{person}{Foyzul Hassan} {and} \bibinfo{person}{Xiaoyin Wang}.} \bibinfo{year}{2018}\natexlab{}.
\newblock \showarticletitle{Hirebuild: An automatic approach to history-driven repair of build scripts}. In \bibinfo{booktitle}{\emph{Proceedings of the 40th international conference on software engineering}}. \bibinfo{pages}{1078--1089}.
\newblock


\bibitem[Henkel et~al\mbox{.}(2020)]%
        {Henkel_2020}
\bibfield{author}{\bibinfo{person}{Jordan Henkel}, \bibinfo{person}{Christian Bird}, \bibinfo{person}{Shuvendu~K. Lahiri}, {and} \bibinfo{person}{Thomas Reps}.} \bibinfo{year}{2020}\natexlab{}.
\newblock \showarticletitle{Learning from, understanding, and supporting DevOps artifacts for docker}. In \bibinfo{booktitle}{\emph{Proceedings of the ACM/IEEE 42nd International Conference on Software Engineering}}. \bibinfo{publisher}{ACM}, \bibinfo{address}{Seoul South Korea}, \bibinfo{pages}{38–49}.
\newblock
\showISBNx{978-1-4503-7121-6}
\urldef\tempurl%
\url{https://doi.org/10.1145/3377811.3380406}
\showDOI{\tempurl}


\bibitem[Hilton et~al\mbox{.}(2016)]%
        {Hilton_2016}
\bibfield{author}{\bibinfo{person}{Michael Hilton}, \bibinfo{person}{Timothy Tunnell}, \bibinfo{person}{Kai Huang}, \bibinfo{person}{Darko Marinov}, {and} \bibinfo{person}{Danny Dig}.} \bibinfo{year}{2016}\natexlab{}.
\newblock \showarticletitle{Usage, costs, and benefits of continuous integration in open-source projects}. In \bibinfo{booktitle}{\emph{Proceedings of the 31st IEEE/ACM International Conference on Automated Software Engineering}}. \bibinfo{publisher}{ACM}, \bibinfo{pages}{426–437}.
\newblock
\showISBNx{978-1-4503-3845-5}
\urldef\tempurl%
\url{https://doi.org/10.1145/2970276.2970358}
\showDOI{\tempurl}


\bibitem[Hora and Valente(2015)]%
        {Hora_2015}
\bibfield{author}{\bibinfo{person}{Andre Hora} {and} \bibinfo{person}{Marco~Tulio Valente}.} \bibinfo{year}{2015}\natexlab{}.
\newblock \showarticletitle{Apiwave: Keeping track of API popularity and migration}. In \bibinfo{booktitle}{\emph{2015 IEEE International Conference on Software Maintenance and Evolution (ICSME)}}. \bibinfo{publisher}{IEEE}, \bibinfo{address}{Bremen, Germany}, \bibinfo{pages}{321–323}.
\newblock
\showISBNx{978-1-4673-7532-0}
\urldef\tempurl%
\url{https://doi.org/10.1109/ICSM.2015.7332478}
\showDOI{\tempurl}


\bibitem[hsweb(2024)]%
        {hsweb-framework_2024}
\bibfield{author}{\bibinfo{person}{hsweb}.} \bibinfo{year}{2024}\natexlab{}.
\newblock
\newblock
\urldef\tempurl%
\url{https://github.com/hs-web/hsweb-framework}
\showURL{%
\tempurl}


\bibitem[Jiao et~al\mbox{.}(2023)]%
        {jiao2023evaluation}
\bibfield{author}{\bibinfo{person}{Mingsheng Jiao}, \bibinfo{person}{Tingrui Yu}, \bibinfo{person}{Xuan Li}, \bibinfo{person}{Guanjie Qiu}, \bibinfo{person}{Xiaodong Gu}, {and} \bibinfo{person}{Beijun Shen}.} \bibinfo{year}{2023}\natexlab{}.
\newblock \showarticletitle{On the Evaluation of Neural Code Translation: Taxonomy and Benchmark}. In \bibinfo{booktitle}{\emph{2023 38th IEEE/ACM International Conference on Automated Software Engineering (ASE)}}. IEEE, \bibinfo{pages}{1529--1541}.
\newblock


\bibitem[Kalliamvakou et~al\mbox{.}(2016)]%
        {Kalliamvakou_2016}
\bibfield{author}{\bibinfo{person}{Eirini Kalliamvakou}, \bibinfo{person}{Georgios Gousios}, \bibinfo{person}{Kelly Blincoe}, \bibinfo{person}{Leif Singer}, \bibinfo{person}{Daniel~M. German}, {and} \bibinfo{person}{Daniela Damian}.} \bibinfo{year}{2016}\natexlab{}.
\newblock \showarticletitle{An in-depth study of the promises and perils of mining GitHub}.
\newblock \bibinfo{journal}{\emph{Empirical Software Engineering}} \bibinfo{volume}{21}, \bibinfo{number}{5} (\bibinfo{date}{Oct} \bibinfo{year}{2016}), \bibinfo{pages}{2035–2071}.
\newblock
\showISSN{1382-3256, 1573-7616}
\urldef\tempurl%
\url{https://doi.org/10.1007/s10664-015-9393-5}
\showDOI{\tempurl}


\bibitem[Kumbhare and Chobe(2014)]%
        {kumbhare2014overview}
\bibfield{author}{\bibinfo{person}{Trupti~A Kumbhare} {and} \bibinfo{person}{Santosh~V Chobe}.} \bibinfo{year}{2014}\natexlab{}.
\newblock \showarticletitle{An overview of association rule mining algorithms}.
\newblock \bibinfo{journal}{\emph{International Journal of Computer Science and Information Technologies}} \bibinfo{volume}{5}, \bibinfo{number}{1} (\bibinfo{year}{2014}), \bibinfo{pages}{927--930}.
\newblock


\bibitem[Li et~al\mbox{.}(2023)]%
        {li2023codeeditor}
\bibfield{author}{\bibinfo{person}{Jia Li}, \bibinfo{person}{Ge Li}, \bibinfo{person}{Zhuo Li}, \bibinfo{person}{Zhi Jin}, \bibinfo{person}{Xing Hu}, \bibinfo{person}{Kechi Zhang}, {and} \bibinfo{person}{Zhiyi Fu}.} \bibinfo{year}{2023}\natexlab{}.
\newblock \showarticletitle{Codeeditor: Learning to edit source code with pre-trained models}.
\newblock \bibinfo{journal}{\emph{ACM Transactions on Software Engineering and Methodology}} \bibinfo{volume}{32}, \bibinfo{number}{6} (\bibinfo{year}{2023}), \bibinfo{pages}{1--22}.
\newblock


\bibitem[Lisowski(2021)]%
        {Lisowski_2021}
\bibfield{author}{\bibinfo{person}{Tomasz Lisowski}.} \bibinfo{year}{2021}\natexlab{}.
\newblock \bibinfo{title}{Top Git Hosting Services for 2022}.
\newblock
\newblock
\urldef\tempurl%
\url{https://gitprotect.io/blog/top-git-hosting-services-for-2022/}
\showURL{%
\tempurl}


\bibitem[Lou et~al\mbox{.}(2019)]%
        {lou2019history}
\bibfield{author}{\bibinfo{person}{Yiling Lou}, \bibinfo{person}{Junjie Chen}, \bibinfo{person}{Lingming Zhang}, \bibinfo{person}{Dan Hao}, {and} \bibinfo{person}{Lu Zhang}.} \bibinfo{year}{2019}\natexlab{}.
\newblock \showarticletitle{History-driven build failure fixing: how far are we?}. In \bibinfo{booktitle}{\emph{Proceedings of the 28th acm sigsoft international symposium on software testing and analysis}}. \bibinfo{pages}{43--54}.
\newblock


\bibitem[Mazrae et~al\mbox{.}(2023)]%
        {Mazrae_2023}
\bibfield{author}{\bibinfo{person}{Pooya~Rostami Mazrae}, \bibinfo{person}{Tom Mens}, \bibinfo{person}{Mehdi Golzadeh}, {and} \bibinfo{person}{Alexandre Decan}.} \bibinfo{year}{2023}\natexlab{}.
\newblock \showarticletitle{On the usage, co-usage and migration of CI/CD tools: A qualitative analysis}.
\newblock \bibinfo{journal}{\emph{Empirical Software Engineering}} \bibinfo{volume}{28}, \bibinfo{number}{2} (\bibinfo{date}{Mar} \bibinfo{year}{2023}), \bibinfo{pages}{52}.
\newblock
\showISSN{1382-3256, 1573-7616}
\urldef\tempurl%
\url{https://doi.org/10.1007/s10664-022-10285-5}
\showDOI{\tempurl}


\bibitem[Mazuran et~al\mbox{.}(2009)]%
        {Mazuran_2009}
\bibfield{author}{\bibinfo{person}{Mirjana Mazuran}, \bibinfo{person}{Elisa Quintarelli}, {and} \bibinfo{person}{Letizia Tanca}.} \bibinfo{year}{2009}\natexlab{}.
\newblock \bibinfo{booktitle}{\emph{Mining Tree-Based Frequent Patterns from XML}}. \bibinfo{series}{Lecture Notes in Computer Science}, Vol.~\bibinfo{volume}{5822}.
\newblock \bibinfo{publisher}{Springer Berlin Heidelberg}, \bibinfo{address}{Berlin, Heidelberg}, \bibinfo{pages}{287–299}.
\newblock
\showISBNx{978-3-642-04956-9}
\urldef\tempurl%
\url{https://doi.org/10.1007/978-3-642-04957-6_25}
\showDOI{\tempurl}


\bibitem[McNicholas et~al\mbox{.}(2008)]%
        {mcnicholas2008}
\bibfield{author}{\bibinfo{person}{Paul~David McNicholas}, \bibinfo{person}{Thomas~Brendan Murphy}, {and} \bibinfo{person}{M O’Regan}.} \bibinfo{year}{2008}\natexlab{}.
\newblock \showarticletitle{Standardising the lift of an association rule}.
\newblock \bibinfo{journal}{\emph{Computational Statistics \& Data Analysis}} \bibinfo{volume}{52}, \bibinfo{number}{10} (\bibinfo{year}{2008}), \bibinfo{pages}{4712--4721}.
\newblock


\bibitem[Meng et~al\mbox{.}(2012)]%
        {Meng2012}
\bibfield{author}{\bibinfo{person}{Sichen Meng}, \bibinfo{person}{Xiaoyin Wang}, \bibinfo{person}{Lu Zhang}, {and} \bibinfo{person}{Hong Mei}.} \bibinfo{year}{2012}\natexlab{}.
\newblock \showarticletitle{A history-based matching approach to identification of framework evolution}. In \bibinfo{booktitle}{\emph{2012 34th International Conference on Software Engineering (ICSE)}}. \bibinfo{pages}{353--363}.
\newblock
\urldef\tempurl%
\url{https://doi.org/10.1109/ICSE.2012.6227179}
\showDOI{\tempurl}


\bibitem[Mossienko(2003)]%
        {Mossienko2003}
\bibfield{author}{\bibinfo{person}{M. Mossienko}.} \bibinfo{year}{2003}\natexlab{}.
\newblock \showarticletitle{Automated Cobol to Java recycling}. In \bibinfo{booktitle}{\emph{Seventh European Conference on Software Maintenance and Reengineering, 2003. Proceedings.}} \bibinfo{pages}{40--50}.
\newblock
\urldef\tempurl%
\url{https://doi.org/10.1109/CSMR.2003.1192409}
\showDOI{\tempurl}


\bibitem[Mukherjee et~al\mbox{.}(2021)]%
        {mukherjee2021fixing}
\bibfield{author}{\bibinfo{person}{Suchita Mukherjee}, \bibinfo{person}{Abigail Almanza}, {and} \bibinfo{person}{Cindy Rubio-Gonz{\'a}lez}.} \bibinfo{year}{2021}\natexlab{}.
\newblock \showarticletitle{Fixing dependency errors for Python build reproducibility}. In \bibinfo{booktitle}{\emph{Proceedings of the 30th ACM SIGSOFT International Symposium on Software Testing and Analysis}}. \bibinfo{pages}{439--451}.
\newblock


\bibitem[Munaiah et~al\mbox{.}(2017)]%
        {Munaiah_2017}
\bibfield{author}{\bibinfo{person}{Nuthan Munaiah}, \bibinfo{person}{Steven Kroh}, \bibinfo{person}{Craig Cabrey}, {and} \bibinfo{person}{Meiyappan Nagappan}.} \bibinfo{year}{2017}\natexlab{}.
\newblock \showarticletitle{Curating GitHub for engineered software projects}.
\newblock \bibinfo{journal}{\emph{Empirical Software Engineering}} \bibinfo{volume}{22}, \bibinfo{number}{6} (\bibinfo{date}{Dec} \bibinfo{year}{2017}), \bibinfo{pages}{3219–3253}.
\newblock
\showISSN{1382-3256, 1573-7616}
\urldef\tempurl%
\url{https://doi.org/10.1007/s10664-017-9512-6}
\showDOI{\tempurl}


\bibitem[Nguyen et~al\mbox{.}(2014)]%
        {nguyen2014statistical}
\bibfield{author}{\bibinfo{person}{Anh~Tuan Nguyen}, \bibinfo{person}{Hoan~Anh Nguyen}, \bibinfo{person}{Tung~Thanh Nguyen}, {and} \bibinfo{person}{Tien~N Nguyen}.} \bibinfo{year}{2014}\natexlab{}.
\newblock \showarticletitle{Statistical learning approach for mining API usage mappings for code migration}. In \bibinfo{booktitle}{\emph{Proceedings of the 29th ACM/IEEE international conference on Automated software engineering}}. \bibinfo{pages}{457--468}.
\newblock


\bibitem[Nguyen et~al\mbox{.}(2016)]%
        {Nguyen2016}
\bibfield{author}{\bibinfo{person}{Anh~Tuan Nguyen}, \bibinfo{person}{Zhaopeng Tu}, {and} \bibinfo{person}{Tien~N. Nguyen}.} \bibinfo{year}{2016}\natexlab{}.
\newblock \showarticletitle{Do Contexts Help in Phrase-Based, Statistical Source Code Migration?}. In \bibinfo{booktitle}{\emph{2016 IEEE International Conference on Software Maintenance and Evolution (ICSME)}}. \bibinfo{pages}{155--165}.
\newblock
\urldef\tempurl%
\url{https://doi.org/10.1109/ICSME.2016.89}
\showDOI{\tempurl}


\bibitem[Nguyen et~al\mbox{.}(2017)]%
        {Nguyen_2017}
\bibfield{author}{\bibinfo{person}{Trong~Duc Nguyen}, \bibinfo{person}{Anh~Tuan Nguyen}, \bibinfo{person}{Hung~Dang Phan}, {and} \bibinfo{person}{Tien~N. Nguyen}.} \bibinfo{year}{2017}\natexlab{}.
\newblock \showarticletitle{Exploring API Embedding for API Usages and Applications}. In \bibinfo{booktitle}{\emph{2017 IEEE/ACM 39th International Conference on Software Engineering (ICSE)}}. \bibinfo{pages}{438--449}.
\newblock
\urldef\tempurl%
\url{https://doi.org/10.1109/ICSE.2017.47}
\showDOI{\tempurl}


\bibitem[Nielsen(1993)]%
        {Response_time}
\bibfield{author}{\bibinfo{person}{Jakob Nielsen}.} \bibinfo{year}{1993}\natexlab{}.
\newblock \bibinfo{title}{Response Time Limits: Article by Jakob Nielsen}.
\newblock
\newblock
\urldef\tempurl%
\url{https://www.nngroup.com/articles/response-times-3-important-limits/}
\showURL{%
\tempurl}


\bibitem[Niemeyer(2023)]%
        {j2swift}
\bibfield{author}{\bibinfo{person}{Pat Niemeyer}.} \bibinfo{year}{2023}\natexlab{}.
\newblock \bibinfo{title}{j2swift}.
\newblock
\newblock
\urldef\tempurl%
\url{https://github.com/patniemeyer/j2swift}
\showURL{%
\tempurl}
\newblock
\shownote{accessed 04-01-2023}.


\bibitem[NLPChina(2024)]%
        {elasticsearch-sql_2024}
\bibfield{author}{\bibinfo{person}{NLPChina}.} \bibinfo{year}{2024}\natexlab{}.
\newblock
\newblock
\urldef\tempurl%
\url{https://github.com/NLPchina/elasticsearch-sql}
\showURL{%
\tempurl}


\bibitem[Pan et~al\mbox{.}(2023)]%
        {pan2023understanding}
\bibfield{author}{\bibinfo{person}{Rangeet Pan}, \bibinfo{person}{Ali~Reza Ibrahimzada}, \bibinfo{person}{Rahul Krishna}, \bibinfo{person}{Divya Sankar}, \bibinfo{person}{Lambert~Pouguem Wassi}, \bibinfo{person}{Michele Merler}, \bibinfo{person}{Boris Sobolev}, \bibinfo{person}{Raju Pavuluri}, \bibinfo{person}{Saurabh Sinha}, {and} \bibinfo{person}{Reyhaneh Jabbarvand}.} \bibinfo{year}{2023}\natexlab{}.
\newblock \showarticletitle{Understanding the effectiveness of large language models in code translation}.
\newblock \bibinfo{journal}{\emph{arXiv preprint arXiv:2308.03109}} (\bibinfo{year}{2023}).
\newblock


\bibitem[Phan et~al\mbox{.}(2017)]%
        {phan_2017}
\bibfield{author}{\bibinfo{person}{Hung~Dang Phan}, \bibinfo{person}{Anh~Tuan Nguyen}, \bibinfo{person}{Trong~Duc Nguyen}, {and} \bibinfo{person}{Tien~N. Nguyen}.} \bibinfo{year}{2017}\natexlab{}.
\newblock \showarticletitle{Statistical Migration of API Usages}. In \bibinfo{booktitle}{\emph{2017 IEEE/ACM 39th International Conference on Software Engineering Companion (ICSE-C)}}. \bibinfo{pages}{47--50}.
\newblock
\urldef\tempurl%
\url{https://doi.org/10.1109/ICSE-C.2017.17}
\showDOI{\tempurl}


\bibitem[Rahman and Parnin(2023)]%
        {Rahman2023}
\bibfield{author}{\bibinfo{person}{Akond Rahman} {and} \bibinfo{person}{Chris Parnin}.} \bibinfo{year}{2023}\natexlab{}.
\newblock \showarticletitle{Detecting and Characterizing Propagation of Security Weaknesses in Puppet-based infrastructure Management}.
\newblock \bibinfo{journal}{\emph{IEEE Transactions on Software Engineering}} (\bibinfo{year}{2023}), \bibinfo{pages}{1--18}.
\newblock
\urldef\tempurl%
\url{https://doi.org/10.1109/TSE.2023.3265962}
\showDOI{\tempurl}


\bibitem[Roziere et~al\mbox{.}(2020)]%
        {Roziere}
\bibfield{author}{\bibinfo{person}{Baptiste Roziere}, \bibinfo{person}{Marie-Anne Lachaux}, \bibinfo{person}{Guillaume Lample}, {and} \bibinfo{person}{Lowik Chanussot}.} \bibinfo{year}{2020}\natexlab{}.
\newblock \showarticletitle{Unsupervised Translation of Programming Languages}.
\newblock \bibinfo{journal}{\emph{NeurIPS 2020}} (\bibinfo{year}{2020}), \bibinfo{pages}{21}.
\newblock


\bibitem[Rzig et~al\mbox{.}(2022b)]%
        {Rzig_2022}
\bibfield{author}{\bibinfo{person}{Dhia~Elhaq Rzig}, \bibinfo{person}{Foyzul Hassan}, \bibinfo{person}{Chetan Bansal}, {and} \bibinfo{person}{Nachiappan Nagappan}.} \bibinfo{year}{2022}\natexlab{b}.
\newblock \showarticletitle{Characterizing the Usage of CI Tools in ML Projects}. In \bibinfo{booktitle}{\emph{Proceedings of the 16th ACM / IEEE International Symposium on Empirical Software Engineering and Measurement}} (Helsinki, Finland) \emph{(\bibinfo{series}{ESEM '22})}. \bibinfo{publisher}{Association for Computing Machinery}, \bibinfo{address}{New York, NY, USA}, \bibinfo{pages}{69–79}.
\newblock
\showISBNx{9781450394277}
\urldef\tempurl%
\url{https://doi.org/10.1145/3544902.3546237}
\showDOI{\tempurl}


\bibitem[Rzig et~al\mbox{.}(2022a)]%
        {Rzig_Hassan_2022}
\bibfield{author}{\bibinfo{person}{Dhia~Elhaq Rzig}, \bibinfo{person}{Foyzul Hassan}, {and} \bibinfo{person}{Marouane Kessentini}.} \bibinfo{year}{2022}\natexlab{a}.
\newblock \showarticletitle{An empirical study on ML DevOps adoption trends, efforts, and benefits analysis}.
\newblock \bibinfo{journal}{\emph{Information and Software Technology}}  \bibinfo{volume}{152} (\bibinfo{date}{Dec} \bibinfo{year}{2022}), \bibinfo{pages}{107037}.
\newblock
\showISSN{0950-5849}
\urldef\tempurl%
\url{https://doi.org/10.1016/j.infsof.2022.107037}
\showDOI{\tempurl}


\bibitem[Salton and McGill(1986)]%
        {cosine_sim}
\bibfield{author}{\bibinfo{person}{Gerard Salton} {and} \bibinfo{person}{Michael~J. McGill}.} \bibinfo{year}{1986}\natexlab{}.
\newblock \bibinfo{booktitle}{\emph{Introduction to Modern Information Retrieval}}.
\newblock \bibinfo{publisher}{McGraw-Hill, Inc.}, \bibinfo{address}{USA}.
\newblock
\showISBNx{0070544840}


\bibitem[Talebipour et~al\mbox{.}(2021)]%
        {Talebipour_2021}
\bibfield{author}{\bibinfo{person}{Saghar Talebipour}, \bibinfo{person}{Yixue Zhao}, \bibinfo{person}{Luka Dojcilović}, \bibinfo{person}{Chenggang Li}, {and} \bibinfo{person}{Nenad Medvidović}.} \bibinfo{year}{2021}\natexlab{}.
\newblock \showarticletitle{UI Test Migration Across Mobile Platforms}. In \bibinfo{booktitle}{\emph{2021 36th IEEE/ACM International Conference on Automated Software Engineering (ASE)}}. \bibinfo{pages}{756--767}.
\newblock
\urldef\tempurl%
\url{https://doi.org/10.1109/ASE51524.2021.9678643}
\showDOI{\tempurl}


\bibitem[Teyton et~al\mbox{.}(2013)]%
        {Teyton_2013}
\bibfield{author}{\bibinfo{person}{Cedric Teyton}, \bibinfo{person}{Jean-Remy Falleri}, {and} \bibinfo{person}{Xavier Blanc}.} \bibinfo{year}{2013}\natexlab{}.
\newblock \showarticletitle{Automatic discovery of function mappings between similar libraries}. In \bibinfo{booktitle}{\emph{2013 20th Working Conference on Reverse Engineering (WCRE)}}. \bibinfo{publisher}{IEEE}, \bibinfo{address}{Koblenz, Germany}, \bibinfo{pages}{192–201}.
\newblock
\showISBNx{978-1-4799-2931-3}
\urldef\tempurl%
\url{https://doi.org/10.1109/WCRE.2013.6671294}
\showDOI{\tempurl}


\bibitem[Travis-CI(2021)]%
        {travis_yaml}
\bibfield{author}{\bibinfo{person}{Travis-CI}.} \bibinfo{year}{2021}\natexlab{}.
\newblock \bibinfo{title}{Travis CI Documentation - Using YAML as a build configuration language}.
\newblock
\newblock
\urldef\tempurl%
\url{https://docs.travis-ci.com/user/build-config-yaml/}
\showURL{%
\tempurl}
\newblock
\shownote{accessed 08-31-2021}.


\bibitem[Travis-CI(2023a)]%
        {os}
\bibfield{author}{\bibinfo{person}{Travis-CI}.} \bibinfo{year}{2023}\natexlab{a}.
\newblock
\newblock
\urldef\tempurl%
\url{https://docs.travis-ci.com/user/multi-os/}
\showURL{%
\tempurl}


\bibitem[Travis-CI(2023b)]%
        {install}
\bibfield{author}{\bibinfo{person}{Travis-CI}.} \bibinfo{year}{2023}\natexlab{b}.
\newblock
\newblock
\urldef\tempurl%
\url{https://docs.travis-ci.com/user/installing-dependencies/}
\showURL{%
\tempurl}


\bibitem[Travis-CI(2023c)]%
        {travis_api}
\bibfield{author}{\bibinfo{person}{Travis-CI}.} \bibinfo{year}{2023}\natexlab{c}.
\newblock \bibinfo{title}{Travis CI Documentation - Triggering Builds}.
\newblock
\newblock
\urldef\tempurl%
\url{https://docs.travis-ci.com/user/triggering-builds/}
\showURL{%
\tempurl}


\bibitem[TravisCI(2022)]%
        {travis}
\bibfield{author}{\bibinfo{person}{TravisCI}.} \bibinfo{year}{2022}\natexlab{}.
\newblock \bibinfo{title}{Home – Travis-CI\_2022}.
\newblock
\newblock
\urldef\tempurl%
\url{https://www.travis-ci.com/}
\showURL{%
\tempurl}


\bibitem[Vassallo et~al\mbox{.}(2019)]%
        {vassallo2019automated}
\bibfield{author}{\bibinfo{person}{Carmine Vassallo}, \bibinfo{person}{Sebastian Proksch}, \bibinfo{person}{Harald~C Gall}, {and} \bibinfo{person}{Massimiliano Di~Penta}.} \bibinfo{year}{2019}\natexlab{}.
\newblock \showarticletitle{Automated reporting of anti-patterns and decay in continuous integration}. In \bibinfo{booktitle}{\emph{2019 IEEE/ACM 41st International Conference on Software Engineering (ICSE)}}. IEEE, \bibinfo{pages}{105--115}.
\newblock


\bibitem[Wang(2019)]%
        {ms_cac}
\bibfield{author}{\bibinfo{person}{Abel Wang}.} \bibinfo{year}{2019}\natexlab{}.
\newblock \bibinfo{title}{What is Configuration as Code? | Microsoft Learn}.
\newblock
\newblock
\urldef\tempurl%
\url{https://learn.microsoft.com/en-us/shows/one-dev-minute/what-is-configuration-as-code--one-dev-question}
\showURL{%
\tempurl}


\bibitem[Wechat-Group(2024)]%
        {Wechat-Group}
\bibfield{author}{\bibinfo{person}{Wechat-Group}.} \bibinfo{year}{2024}\natexlab{}.
\newblock \bibinfo{title}{Wechat-Group/WxJava}.
\newblock
\newblock
\urldef\tempurl%
\url{https://github.com/Wechat-Group/WxJava}
\showURL{%
\tempurl}


\bibitem[Widder et~al\mbox{.}(2018)]%
        {widder2018ci}
\bibfield{author}{\bibinfo{person}{David Widder}, \bibinfo{person}{Michael Hilton}, \bibinfo{person}{Christian K\"{a}stner}, {and} \bibinfo{person}{Bogdan Vasilescu}.} \bibinfo{year}{2018}\natexlab{}.
\newblock \showarticletitle{I’m Leaving You, Travis: A Continuous Integration Breakup Story}. In \bibinfo{booktitle}{\emph{International Conference on Mining Software Repositories}} \emph{(\bibinfo{series}{MSR})}. \bibinfo{publisher}{ACM}, \bibinfo{pages}{165--169}.
\newblock
\urldef\tempurl%
\url{https://doi.org/10.1145/3196398.3196422}
\showDOI{\tempurl}


\bibitem[Xie et~al\mbox{.}(2020)]%
        {Xie_2020}
\bibfield{author}{\bibinfo{person}{Chunli Xie}, \bibinfo{person}{Xia Wang}, \bibinfo{person}{Cheng Qian}, {and} \bibinfo{person}{Mengqi Wang}.} \bibinfo{year}{2020}\natexlab{}.
\newblock \showarticletitle{A Source Code Similarity Based on Siamese Neural Network}.
\newblock \bibinfo{journal}{\emph{Applied Sciences}} \bibinfo{volume}{10}, \bibinfo{number}{21} (\bibinfo{date}{Oct} \bibinfo{year}{2020}), \bibinfo{pages}{7519}.
\newblock
\showISSN{2076-3417}
\urldef\tempurl%
\url{https://doi.org/10.3390/app10217519}
\showDOI{\tempurl}


\bibitem[Xu et~al\mbox{.}(2020)]%
        {xu2020sledge}
\bibfield{author}{\bibinfo{person}{Bo Xu}, \bibinfo{person}{Song Wu}, \bibinfo{person}{Jiang Xiao}, \bibinfo{person}{Hai Jin}, \bibinfo{person}{Yingxi Zhang}, \bibinfo{person}{Guoqiang Shi}, \bibinfo{person}{Tingyu Lin}, \bibinfo{person}{Jia Rao}, \bibinfo{person}{Li Yi}, {and} \bibinfo{person}{Jizhong Jiang}.} \bibinfo{year}{2020}\natexlab{}.
\newblock \showarticletitle{Sledge: Towards efficient live migration of docker containers}. In \bibinfo{booktitle}{\emph{2020 IEEE 13th International Conference on Cloud Computing (CLOUD)}}. IEEE, \bibinfo{pages}{321--328}.
\newblock


\bibitem[Yang et~al\mbox{.}(2023)]%
        {yang2023syntax}
\bibfield{author}{\bibinfo{person}{Guang Yang}, \bibinfo{person}{Yu Zhou}, \bibinfo{person}{Xiang Chen}, \bibinfo{person}{Xiangyu Zhang}, \bibinfo{person}{Yiran Xu}, \bibinfo{person}{Tingting Han}, {and} \bibinfo{person}{Taolue Chen}.} \bibinfo{year}{2023}\natexlab{}.
\newblock \showarticletitle{A Syntax-Guided Multi-Task Learning Approach for Turducken-Style Code Generation}.
\newblock \bibinfo{journal}{\emph{arXiv preprint arXiv:2303.05061}} (\bibinfo{year}{2023}).
\newblock


\bibitem[Zampetti et~al\mbox{.}(2019)]%
        {zampetti2019study}
\bibfield{author}{\bibinfo{person}{Fiorella Zampetti}, \bibinfo{person}{Gabriele Bavota}, \bibinfo{person}{Gerardo Canfora}, {and} \bibinfo{person}{Massimiliano Di~Penta}.} \bibinfo{year}{2019}\natexlab{}.
\newblock \showarticletitle{A study on the interplay between pull request review and continuous integration builds}. In \bibinfo{booktitle}{\emph{2019 IEEE 26th international conference on software analysis, evolution and reengineering (SANER)}}. IEEE, \bibinfo{pages}{38--48}.
\newblock


\bibitem[Zhang et~al\mbox{.}(2018)]%
        {zhang2018cd}
\bibfield{author}{\bibinfo{person}{Yang Zhang}, \bibinfo{person}{Bogdan Vasilescu}, \bibinfo{person}{Huaimin Wang}, {and} \bibinfo{person}{Vladimir Filkov}.} \bibinfo{year}{2018}\natexlab{}.
\newblock \showarticletitle{One Size Does Not Fit All: An Empirical Study of Containerized Continuous Deployment Workflows}. In \bibinfo{booktitle}{\emph{Joint European Software Engineering Conference and Symposium on the Foundations of Software Engineering}} \emph{(\bibinfo{series}{ESEC/FSE})}. \bibinfo{publisher}{ACM}, \bibinfo{pages}{295--306}.
\newblock
\urldef\tempurl%
\url{https://doi.org/10.1145/3236024.3236033}
\showDOI{\tempurl}


\bibitem[Zhao et~al\mbox{.}(2017)]%
        {zhao2017impact}
\bibfield{author}{\bibinfo{person}{Yangyang Zhao}, \bibinfo{person}{Alexander Serebrenik}, \bibinfo{person}{Yuming Zhou}, \bibinfo{person}{Vladimir Filkov}, {and} \bibinfo{person}{Bogdan Vasilescu}.} \bibinfo{year}{2017}\natexlab{}.
\newblock \showarticletitle{The impact of continuous integration on other software development practices: a large-scale empirical study}. In \bibinfo{booktitle}{\emph{2017 32nd IEEE/ACM International Conference on Automated Software Engineering (ASE)}}. IEEE, \bibinfo{pages}{60--71}.
\newblock


\bibitem[Zhong et~al\mbox{.}(2010)]%
        {Zhong2010}
\bibfield{author}{\bibinfo{person}{Hao Zhong}, \bibinfo{person}{Suresh Thummalapenta}, \bibinfo{person}{Tao Xie}, \bibinfo{person}{Lu Zhang}, {and} \bibinfo{person}{Qing Wang}.} \bibinfo{year}{2010}\natexlab{}.
\newblock \showarticletitle{Mining API mapping for language migration}. In \bibinfo{booktitle}{\emph{2010 ACM/IEEE 32nd International Conference on Software Engineering}}, Vol.~\bibinfo{volume}{1}. \bibinfo{pages}{195--204}.
\newblock
\urldef\tempurl%
\url{https://doi.org/10.1145/1806799.1806831}
\showDOI{\tempurl}


\end{thebibliography}

\end{document}